\documentclass[fleqn,aps,prl,nofootinbib,superscriptaddress,letterpaper]{revtex4}
\usepackage[hyperindex,breaklinks]{hyperref}
\usepackage{graphicx} 
\usepackage{dcolumn}  
\usepackage{bm}       
\usepackage{feynmf}
\usepackage{slashed}
\usepackage{multirow}
\usepackage{amsmath}
\usepackage{amssymb}
\usepackage{color}

\begin{document}

\title{Dark matter direct search sensitivity of the PandaX-4T experiment}
\author{Hongguang ZHANG}
\affiliation{INPAC and School of Physics and Astronomy, Shanghai Jiao
  Tong University, \\
  Shanghai Laboratory for Particle Physics and
  Cosmology, Shanghai 200240, China}
\author{Abdusalam ABDUKERIM}
\affiliation{School of Physics and Technology, Xinjiang University, Urumqi 830046, China}
\author{Xun CHEN}
\affiliation{INPAC and School of Physics and Astronomy, Shanghai Jiao
  Tong University, \\
  Shanghai Laboratory for Particle Physics and
  Cosmology, Shanghai 200240, China}
\author{Yunhua CHEN} 
\affiliation{Yalong River
  Hydropower Development Company, Ltd., 288 Shuanglin Road, Chengdu
  610051, China} 
\author{Xiangyi CUI}
\author{Binbin DONG}
\affiliation{INPAC and School of Physics and Astronomy, Shanghai Jiao
  Tong University, \\
  Shanghai Laboratory for Particle Physics and
  Cosmology, Shanghai 200240, China}
\author{Deqing FANG}
\affiliation{Shanghai Institute of Applied Physics,
  Chinese Academy of Sciences, Shanghai 201800, China}
\author{Changbo FU}
\author{Karl GIBONI}
\author{Franco GIULIANI}
\author{Linhui GU}
\affiliation{INPAC and School of Physics and Astronomy, Shanghai Jiao
  Tong University, \\
  Shanghai Laboratory for Particle Physics and
  Cosmology, Shanghai 200240, China}
\author{Xuyuan GUO}
\affiliation{Yalong River
  Hydropower Development Company, Ltd., 288 Shuanglin Road, Chengdu
  610051, China} 
\author{Zhifan GUO}
\affiliation{School of Mechanical Engineering, Shanghai Jiao Tong
  University, Shanghai 200240, China} 
\author{Ke HAN}
\author{Changda HE}
\affiliation{INPAC and School of Physics and Astronomy, Shanghai Jiao
  Tong University, \\
  Shanghai Laboratory for Particle Physics and
  Cosmology, Shanghai 200240, China}
\author{Shengming HE}
\affiliation{Yalong River Hydropower Development Company, Ltd., 288 Shuanglin Road, Chengdu 610051, China} 
\author{Di HUANG}
\affiliation{INPAC and School of Physics and Astronomy, Shanghai Jiao
  Tong University, \\
  Shanghai Laboratory for Particle Physics and
  Cosmology, Shanghai 200240, China}
\author{Xingtao HUANG}
\affiliation{School of Physics and Key Laboratory of Particle Physics and Particle Irradiation
  (MOE), Shandong University, Jinan 250100, China}
\author{Zhou HUANG}{}
\affiliation{INPAC and School of Physics and Astronomy, Shanghai Jiao
  Tong University, \\
  Shanghai Laboratory for Particle Physics and
  Cosmology, Shanghai 200240, China}
\author{Peng JI}
\affiliation{School of Physics, Nankai University, Tianjin 300071, China}
\author{Xiangdong JI}
\email[Spokesperson, Corresponding author: ]{xdji@sjtu.edu.cn}
\affiliation{INPAC and School of Physics and Astronomy, Shanghai Jiao
  Tong University, \\
  Shanghai Laboratory for Particle Physics and
  Cosmology, Shanghai 200240, China}
\affiliation{Tsung-Dao Lee Institute, Shanghai 200240, China}
\author{Yonglin JU}
\affiliation{School of Mechanical Engineering, Shanghai Jiao Tong University, Shanghai 200240, China} 
\author{Shaoli LI}
\author{Yao LI}
\author{Heng LIN}
\affiliation{INPAC and School of Physics and Astronomy, Shanghai Jiao
  Tong University, \\
  Shanghai Laboratory for Particle Physics and
  Cosmology, Shanghai 200240, China}
\author{Huaxuan LIU}
\affiliation{School of Mechanical Engineering, Shanghai Jiao Tong University, Shanghai 200240, China} 
\author{Jianglai LIU}
\affiliation{INPAC and School of Physics and Astronomy, Shanghai Jiao Tong University, \\
  Shanghai Laboratory for Particle Physics and Cosmology, Shanghai 200240, China}
\affiliation{Tsung-Dao Lee Institute, Shanghai 200240, China}
\author{Yugang MA}
\affiliation{Shanghai Institute of Applied Physics, Chinese Academy of Sciences, Shanghai 201800, China}
\author{Yajun MAO}
\affiliation{School of Physics, Peking University, Beijing 100871,China} 
\author{Kaixiang NI}
\affiliation{INPAC and School of Physics and Astronomy, Shanghai Jiao Tong University, \\
  Shanghai Laboratory for Particle Physics and Cosmology, Shanghai 200240, China}
\author{Jinhua NING}
\affiliation{Yalong River Hydropower Development Company, Ltd., 288 Shuanglin Road, Chengdu 610051, China} 
\author{Xiangxiang REN}
\author{Fang SHI}
\affiliation{INPAC and School of Physics and Astronomy, Shanghai Jiao Tong University, \\
  Shanghai Laboratory for Particle Physics and Cosmology, Shanghai 200240, China}
\author{Andi TAN}
\affiliation{Department of Physics, University of Maryland, College Park, Maryland 20742, USA} 
\author{Anqing WANG}
\affiliation{School of Physics and Key Laboratory of Particle Physics and Particle Irradiation
  (MOE), Shandong University, Jinan 250100, China}
\author{Cheng WANG}
\affiliation{School of Mechanical Engineering, Shanghai Jiao Tong University, Shanghai 200240, China} 
\author{Hongwei WANG}
\affiliation{Shanghai Institute of Applied Physics,
  Chinese Academy of Sciences, Shanghai 201800, China}
\author{Meng WANG}
\affiliation{School of Physics and Key Laboratory of Particle Physics and Particle Irradiation
  (MOE), Shandong University, Jinan 250100, China}
\author{Qiuhong WANG}
\affiliation{Shanghai Institute of Applied Physics,
  Chinese Academy of Sciences, Shanghai 201800, China}
\author{Siguang WANG}
\affiliation{School of Physics, Peking University, Beijing 100871,China} 
\author{Xiuli WANG}
\affiliation{School of Mechanical Engineering, Shanghai Jiao Tong University, Shanghai 200240, China} 
\author{Xuming WANG}
\affiliation{INPAC and School of Physics and Astronomy, Shanghai Jiao Tong University, \\
  Shanghai Laboratory for Particle Physics and Cosmology, Shanghai 200240, China}
\author{Zhou WANG}
\affiliation{School of Mechanical Engineering, Shanghai Jiao Tong University, Shanghai 200240, China} 
\author{Mengmeng WU}
\affiliation{School of Physics, Nankai University, Tianjin 300071, China}
\author{Shiyong WU}
\affiliation{Yalong River Hydropower Development Company, Ltd., 288 Shuanglin Road, Chengdu 610051, China} 
\author{Jingkai XIA}
\affiliation{INPAC and School of Physics and Astronomy, Shanghai Jiao Tong University, \\
  Shanghai Laboratory for Particle Physics and Cosmology, Shanghai 200240, China}
\author{Mengjiao XIAO}
\affiliation{Department of Physics, University of Maryland, College Park, Maryland 20742, USA} 
\affiliation{Center of High Energy Physics, Peking University, Beijing 100871, China}
\author{Pengwei XIE}
\affiliation{Tsung-Dao Lee Institute, Shanghai 200240, China}
\author{Binbin YAN}
\affiliation{School of Physics and Key Laboratory of Particle Physics and Particle Irradiation
  (MOE), Shandong University, Jinan 250100, China}
\author{Jijun YANG}
\author{Yong YANG}
\affiliation{INPAC and School of Physics and Astronomy, Shanghai Jiao Tong University, \\
  Shanghai Laboratory for Particle Physics and Cosmology, Shanghai 200240, China}
\author{Chunxu YU}
\affiliation{School of Physics, Nankai University, Tianjin 300071, China}
\author{Jumin YUAN}
\affiliation{School of Physics and Key Laboratory of Particle Physics and Particle Irradiation
  (MOE), Shandong University, Jinan 250100, China}
\author{Jianfeng YUE}
\affiliation{Yalong River Hydropower Development Company, Ltd., 288 Shuanglin Road, Chengdu 610051, China} 
\author{Dan ZHANG}
\affiliation{Department of Physics, University of Maryland, College Park, Maryland 20742, USA} 
\author{Tao ZHANG}
\author{Li ZHAO}
\affiliation{INPAC and School of Physics and Astronomy, Shanghai Jiao Tong University, \\
  Shanghai Laboratory for Particle Physics and Cosmology, Shanghai 200240, China}
\author{Jifang ZHOU}
\affiliation{Yalong River Hydropower Development Company, Ltd., 288 Shuanglin Road, Chengdu 610051, China} 
\author{Ning ZHOU}
\email[Corresponding author: ]{nzhou@sjtu.edu.cn}
\affiliation{INPAC and School of Physics and Astronomy, Shanghai Jiao Tong University, \\
  Shanghai Laboratory for Particle Physics and Cosmology, Shanghai 200240, China}
\author{Xiaopeng ZHOU}
\affiliation{School of Physics, Peking University, Beijing 100871,China}

\begin{abstract}
  The PandaX-4T experiment, a four-ton scale dark matter
  direct detection experiment, is being planned at the China Jinping
  Underground Laboratory.  In this paper we present a simulation study
  of the expected background in this experiment. In a 2.8-ton fiducial
  mass and the signal region between 1 to 10~keV electron equivalent
  energy, the total electron recoil background is found to be
  $\rm 4.9\cdot 10^{-5}~(kg\cdot day \cdot keV)^{-1}$. The nuclear recoil
  background in the same region is $\rm 2.8\cdot 10^{-7}~(kg \cdot day
  \cdot keV)^{-1}$. With an exposure of 5.6 ton-years, the sensitivity
  of PandaX-4T could reach a minimum spin-independent dark
  matter-nucleon cross section of $\rm 6\cdot 10^{-48}~cm^2$ at a dark matter mass of 40~GeV/$c^2$.
\end{abstract}

\maketitle

\section{Introduction}
Precise astrophysical and cosmological observations indicate that most
of the universe is composed of the dark matter and dark
energy~\cite{dm1,dm2}. In the standard model of cosmology, the dark
matter contributes up to $26.8\%$ of the total content of the
universe~\cite{dm3}.  Near our solar system, the local dark matter
density is estimated to be 0.3 $\rm GeV/cm^{3}$~\cite{density}.  One of
the most favored dark matter particle candidates are the so-called
Weakly Interacting Massive Particles (WIMPs) predicted in many
theories beyond the Standard Model of particle
physics~\cite{wimp2}. Theories also predict a feeble interaction
between the WIMPs and the normal matter, so when WIMPs pass through a
detector on Earth, they could produce detectable low-energy nuclear
recoil signals. This is the principle of dark matter direct
detection~\cite{wimp3}.

Dual-phase xenon detectors are currently leading in the search for
WIMPs with mass ranging from a few GeV/$c^2$ to several
TeV/$c^2$~\cite{ddreview}. In a dual-phase xenon detector, the
WIMP-induced nuclear recoil (NR) events have two signals: a prompt $S1$
signal due to scintillation photons in the liquid xenon and a delayed
$S2$ signal from ionization electrons, which drift to the liquid
surface and then produce electroluminescence
photons in the xenon gas. The $S1$ and $S2$ signals are collected by
two arrays of photomultiplier tubes (PMTs) on the top and bottom of
the detector. The WIMP-nucleon scattering vertex in three-dimensions
can be reconstructed based on the photon pattern on the PMT arrays and
the time difference between the $S1$ and $S2$ signals. This type of
detector is also commonly referred to as the time projection chamber (TPC). Most
of the background events in the detector are from electron recoils
(ER). These events have different proportion in $S1$ and $S2$
signals compared to the NR events, so a large fraction can be
identified and removed.  During the past years, dual phase xenon
experiments LUX\cite{lux2017}, XENON1T\cite{xenon2017} and PandaX-II~\cite{tan2016,cui2017} have been
producing new data continuously and pushed the constraints on the
WIMP-nucleon scattering cross section down to a lowest level of $\rm
10^{-46}cm^2$ at a dark matter mass around 50 GeV/$c^2$.  In this paper, we shall discuss the
upgrade of PandaX-II, the 4-ton scale PandaX-4T experiment, and
evaluate its potential sensitivity to WIMP detection. 

The rest of this paper is organized as follows. Sec.~\ref{4Tdetector}
describes the PandaX-4T experiment and detector design.
The ER and NR backgrounds are discussed in Sec.~\ref{er-background} and
Sec.~\ref{nr-background}, respectively, followed by a summary in Sec.~\ref{background-summary}.
The WIMP sensitivity is presented in Sec.~\ref{physics},
followed by a conclusion in Sec.~\ref{conclusion}.

\section{The PandaX-4T experiment}
\label{4Tdetector}

\begin{figure}[thbp]
  \centering
  \includegraphics[width=4cm]{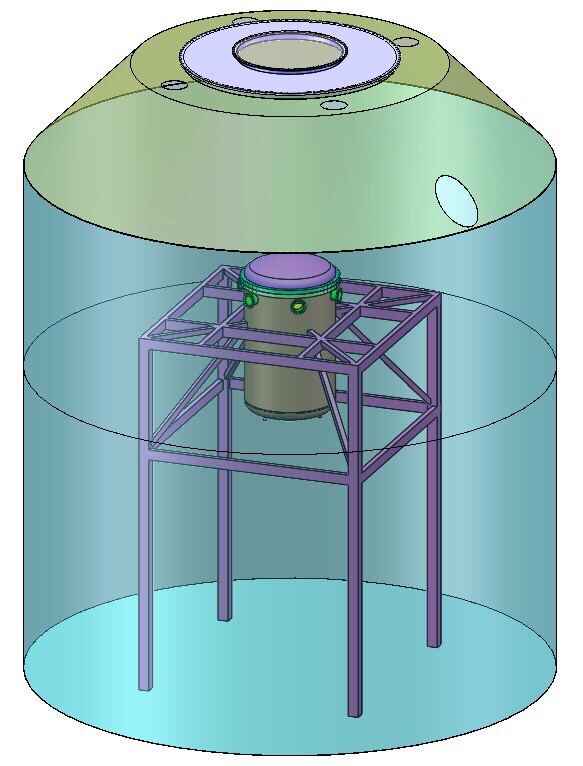}
  \includegraphics[width=3.5cm]{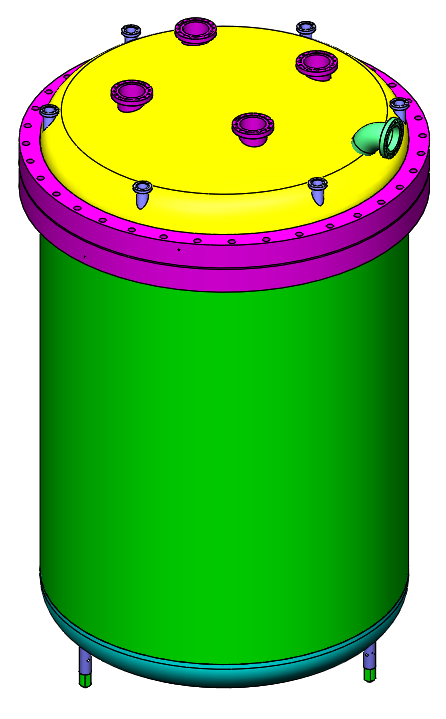}
  \caption{Left: The ultrapure water shield for PandaX-4T. Right: the inner cryostat. }
  \label{vessel} 
\end{figure}

The PandaX-4T experiment will be located in the B2 experimental hall
in the second phase of China Jinping underground Laboratory
(CJPL-II). With a 2400-m marble overburden, the cosmic ray flux in
CJPL is reduced to a negligible level ($\rm 2\cdot 10^{-10} cm^{-2}s^{-1}$). In addition, the PandaX-4T
detector will be placed in the center of 13~m tall, 10~m diameter
water tank (Fig.~\ref{vessel}) containing ultrapure water to suppress
the external gammas and neutrons from the laboratory environment to a
negligible level. 

To ensure an ultra-clean background for the PandaX-4T experiment,
materials with a low level of radioactivity are selected for the
detector construction. The central cylindrical TPC has a transverse
diameter of 1.2~m and a height of 1.3~m.  It is contained inside a
cryostat with an outer vessel and inner vessel constructed with low
background stainless steel (SS).  The inner vessel which holds the
liquid xenon ($-$100 C) is composed of a barrel and two domes. The
barrel is approximately 1.8~meters in height and 1.3~meter in
diameter, with a weight of about 0.5~ton. The bottom dome is welded to
the barrel. The top dome has a SS flange which seals to the top of the
barrel, with a total weight of 0.5~ton.  The outer vessel, with a gap
of 12.5~cm to the inner vessel, provides a vacuum thermal insulation
for the inner vessel, and the total weight is approximately
1.5~tons. The total mass of liquid xenon is approximately 6~tons, and
the liquid xenon contained in the sensitive volume (TPC) is 4~tons
(see Figure~\ref{vessel}).

\begin{figure}[htbp]
  \centering
  \includegraphics[width=3.8cm]{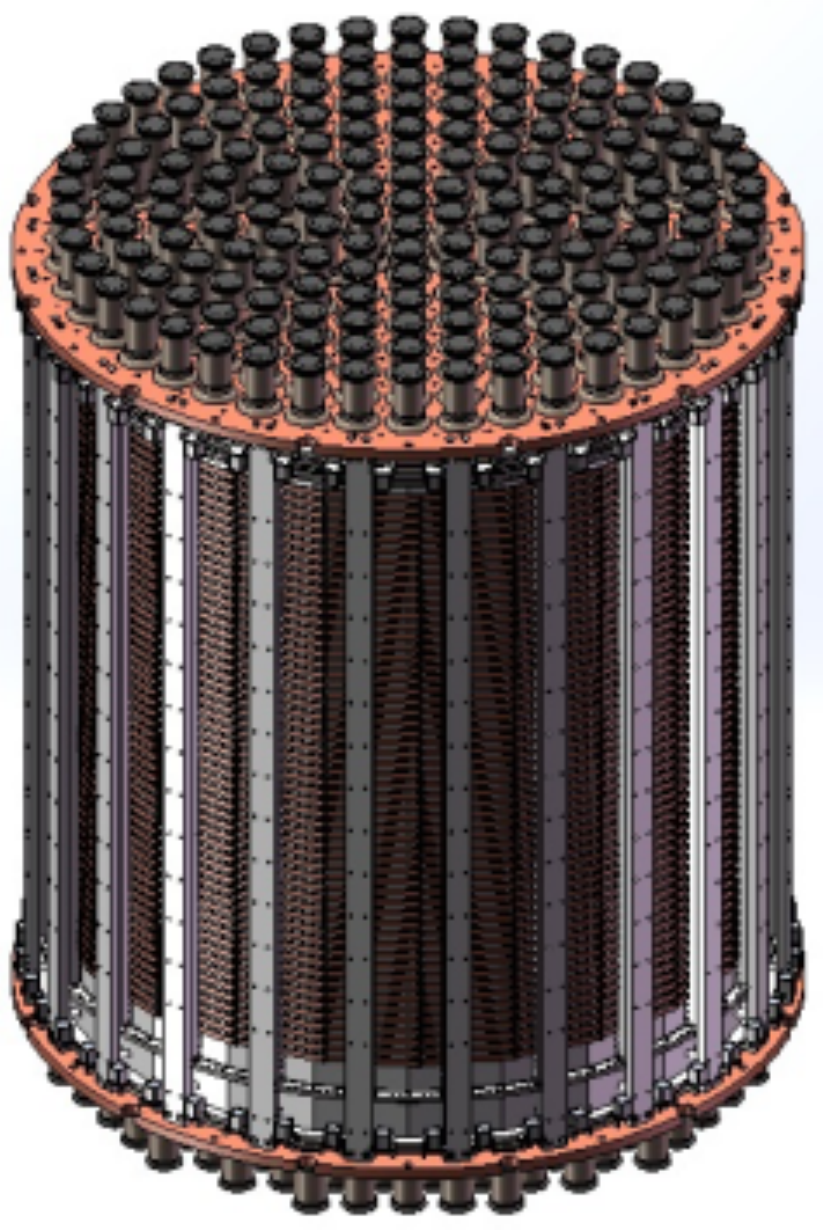}
  \includegraphics[width=4cm]{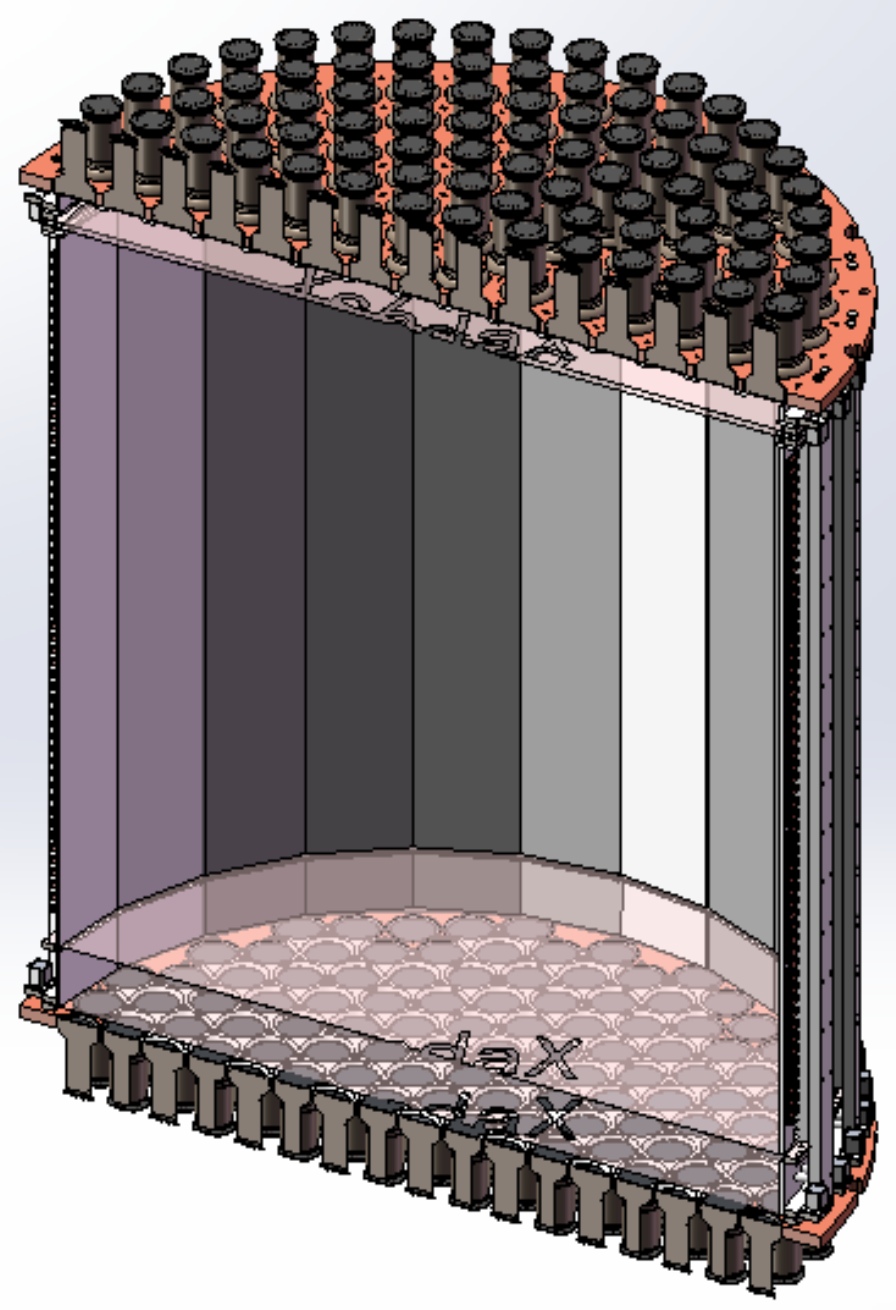}
  \caption{The diagram of the PandaX-4T TPC.}
  \label{tpc} 
\end{figure}

The TPC consists of two 20~mm thick copper plates and 24 pieces of
6~mm thick interlocking polytetrafluoroethylene (PTFE) panels, which
laterally form an approximately cylindrical structure with a diameter
of 1.2~meters, as shown in Figure~\ref{tpc}.  Four highly transparent
electrodes (from top to bottom: anode, gate, cathode and screening)
are designed inside the TPC to produce needed electric fields. The
electrodes are made of SS rings with 0.2~mm SS mesh fixed on them. The
cathode is placed 100~mm above the bottom copper plate, and the gate
and anode electrodes are separated by 10 mm with liquid surface in the
middle.  The drift distance between the gate and cathode electrodes is
1.2 meters. The design drift field is 400~V/cm in the liquid xenon,
and the amplification field in the gas xenon is 6000~V/cm.
Directly on top of the bottom copper plate, there is a screening
electrode which is grounded. In addition, approximately 60 copper
shaping rings surrounding the TPC are designed to maintain an uniform
vertical drift field inside the TPC, especially close to the PTFE
wall. The total mass of the shaping rings is approximately 45~kg. 24
pieces of PTFE pillars are designed to connect the top and bottom
copper plates and to hold the electrodes and shaping rings.  A layer
of PTFE is attached to the inner wall of the cryostat, and the liquid
xenon between the TPC and the wall is a highly reflective and
optically separated region used as a veto compartment.

For the signal photon collection, the low background 3-inch PMTs
Hamamatsu R11410-23 are fixed onto the top and bottom copper plates
with photocathodes facing inward.
There are 169 top PMTs placed in a concentric circular pattern,
approximately $50$~mm above the liquid surface, and 199 bottom PMTs in
a compact hexagonal structure placed approximately 1.3~meter below the
liquid surface. The space between the PMT photocathodes is covered
with PTFE for better photon collection.  Two circles of
1-inch Hamamatsu R8520-406 PMTs are placed at the top and bottom of
the veto compartment to collect photons produced by peripheral
background.

\section{Background estimation}
\label{simulation}
The WIMP signals are expected to be single scatterering events,
uniformly distributed in the target volume.  WIMP-nucleus recoil
signals yield a typical energy of tens of keV, and a nominal WIMP
search window is between 1 to 10 keV electron equivalent
energy. Due to the self-shielding from the liquid xenon, the
radioactive background from the detector materials will be located
close to the edge of the target, and a large fraction can also be
removed due to multiple scattering vertices. The other backgrounds due
to trace amount of radioactive noble gases, e.g. $\rm ^{85}Kr$, $\rm
^{222}Rn$ and $\rm ^{136}Xe$, would diffuse uniformly in the target,
so would the events due to solar neutrinos.

The radioactivities of all materials used in the detector construction
were measured in a high purity Germanium (HPGe) gamma-ray detector
located at CJPL, and are inputs to the background estimation. 
For the PTFE materials, since the natural radioactivities are too low to be measured
by HPGe, we adopted values from XENON1T~\cite{xenon1t} which were measured
after neutron activation. For other materials, when only upper
limits are available from the measurement, we take them as the
central values for conservative background estimation. The radioactivities of the PMT were
measured by us as a single component. For the ER background estimation
for the PMTs, we took the measured values or upper limits and assumed
that they originate from the quartz windows. But the corresponding NR
background was computed component by component (e.g. quartz window,
ceramic stem) based on the measured values from XENON1T.
The radioactivity of the materials is summarized in Table~\ref{radioactivity}.

\begin{table*}[t]
  \footnotesize
  \caption{Material Radioactivity for PandaX-4T: The radioactivity values of PTFE and R11410 PMT components (window and stem) are took from the Xenon1T experiment, while others are the same as the PandaX-II experiment.}
  \label{radioactivity}
  \begin{tabular*}{\textwidth}{@{\extracolsep{\fill}}lcccccccc}
    \toprule \multicolumn{9}{c}{\bf Radioactivity: mBq/unit} \\\hline
    Material & $^{60}\rm{Co}$ & $^{40}\rm{K}$ & $^{137}\rm{Cs}$ & $^{238}\rm{U}_e$ & $^{238}\rm{U}_l$ & $^{235}\rm{U}$ & $^{232}\rm{Th}_e$ & $^{232}\rm{Th}_l$ \\\hline
    SS 3000~kg &	1.03 &	13.95 & 2.36 &	1.7  & 1.7  & 2.43 & 2.74 & 2.74 \\\hline
    Copper 200~kg       &0.2 &	4     & 0.16 &	0.38 & 0.38 & 0.86 & 0.51 & 0.51 \\\hline
    PTFE 200~kg (XENON1T)&	0.027 &	0.34 &	0.17 &	0.25 & 0.12 & 0.01 & 0.04 & 0.07 \\\hline
    R11410 PMT 368~pics &3.5 &13 & 0.3   &	0.94 & 0.94 & 1.17 & 1.6  & 1.6  \\\hline
    R11410 Window(Quartz) 368~pics (XENON1T)& 0.01 & 0.02 & 0.01 &  1.2 & 0.07 & 0.02 & 0.03 & 0.03 \\\hline
    R11410 Stem($\rm Al_2 O_3$) 368~pics (XENON1T)&	0.02 &	1.1 & 0.02 & 2.4 & 0.26 & 0.11 & 0.23 & 0.11 \\\hline
    R8520 PMT 144~pics& 0.75 &	8.15 & 0.17 & 0.25 & 0.25 & 0.11  & 0.46 & 0.46 \\\hline
    %\bottomrule
  \end{tabular*}
  
\end{table*}

A customized Geant4~\cite{geant4} (ver. 10.2p02) simulation program
with the above geometry implemented, BambooMC, is used to
simulate the decay of the radioactive isotopes and their energy
depositions inside the liquid xenon. For each deposition, information
of the deposited energy, position, time and interaction type are
recorded. A post-simulation reconstruction algorithm is developed to
cluster the neighboring depositions into individually measurable scatters so that the
comparison between the MC simulation and data is possible. Only
scatters with an energy exceeding the threshold of PandaX-II ($\rm 1~keV$ electron equivalent energy) are considered.

\subsection{ER Background}
\label{er-background}
All sources of ER background are considered in this section, which
include the radioactive contamination from detector materials,
intrinsic radioactive isotopes such as $^{85}\rm{Kr}$,
$^{222}\rm{Rn}$, and $^{136}\rm{Xe}$ in liquid xenon, and neutrinos
from the sun. The rate of the background is calculated between 1 to 10
keV$_{ee}$, with a 2.8-ton fiducial cut.

As the detector is shielded by the ultrapure water shield, the
laboratory background from the concrete wall, water tank, and others
materials in the laboratory is suppressed to a negligible level.

\begin{figure}[htbp]
  \centering
  \includegraphics[width=8cm]{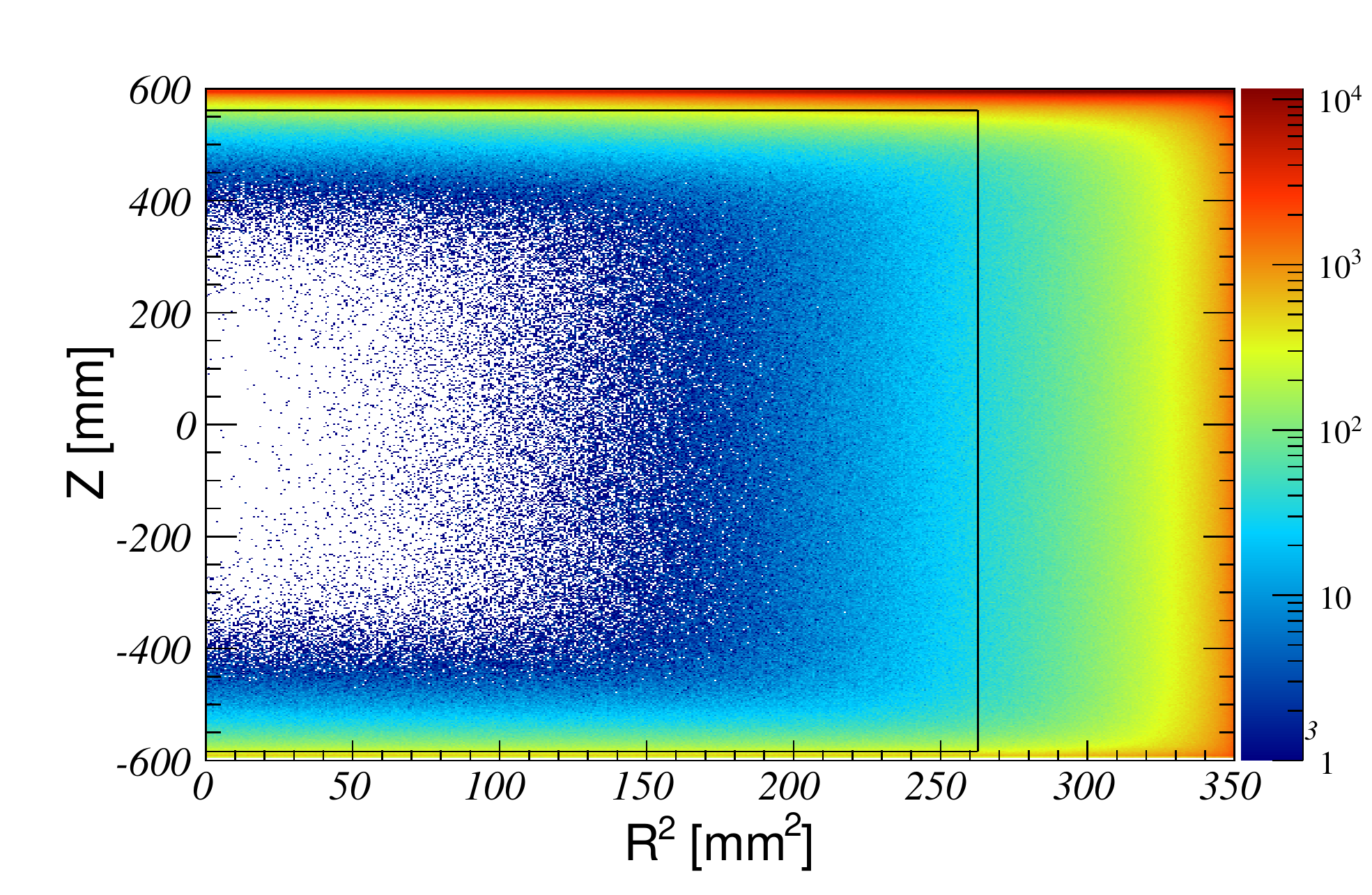}
  \caption{The distribution of the simulated ER background events
    after the dark matter signal selection cuts in the TPC.}
  \label{er-dist} 
\end{figure}

In the detector simulation, the materials we considered include the
SS, copper, PTFE and the PMTs using the radioactivity levels from
Table~\ref{radioactivity}. For each isotope, $10^8$ decays are
generated uniformly in the component volume. Within the dark matter
signal selection windows (energy window, fiducial volume and single
scattering), we count the survival events and convert to the
background rate. The ER background distribution inside the TPC is
shown in Figure~\ref{er-dist}.

For the ER background, some single scatter events in the TPC may also
deposit energy in the veto compartment, and the veto PMTs provide
additional background suppression ability. We assume that the veto
compartment triggers at an energy deposition larger than 150 keV, a
conservative value adopted from PandaX-II experiment. When such a veto 
is applied to the ER background,  we find approximately 50\% of the ER
background from the TPC materials can be rejected. 

\begin{figure}[htbp]
  \centering
  \includegraphics[width=8cm]{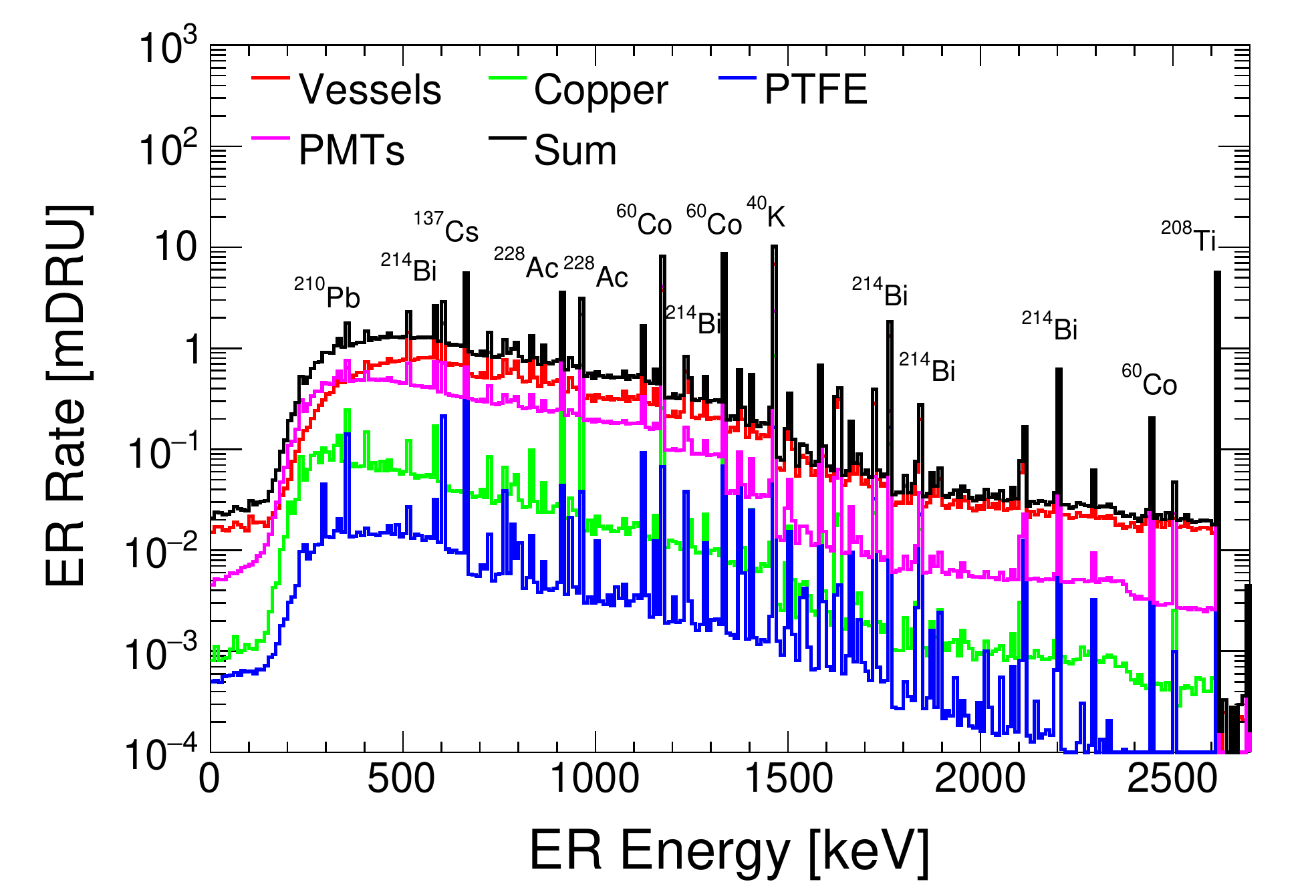}
  \caption{The simulated ER energy spectra from detector materials and
    various components (PTFE, PMTs, SS vessels and copper).}
  \label{er-spectrum} 
\end{figure}

\begin{figure}[htbp]
  \centering
  \includegraphics[width=7cm]{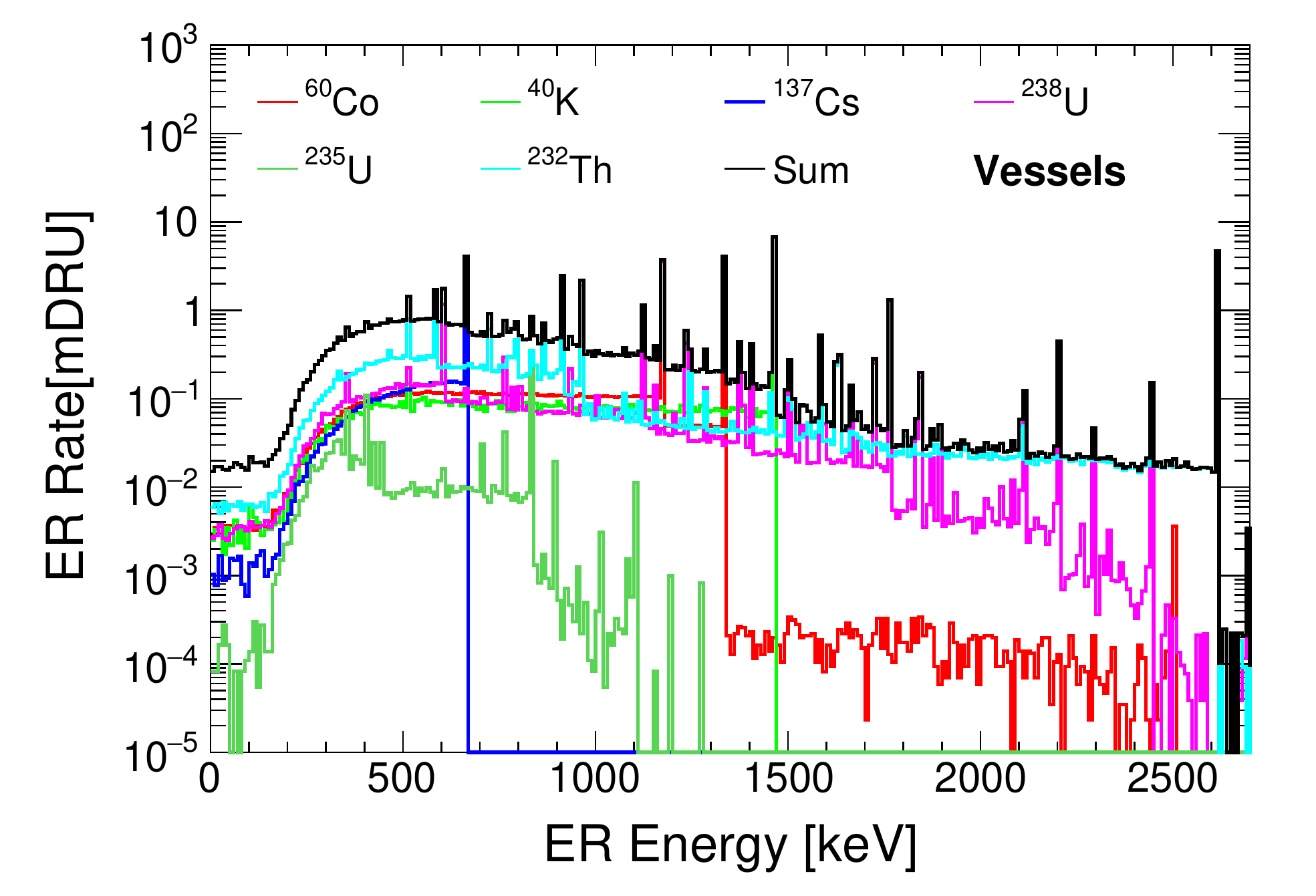}
  \includegraphics[width=7cm]{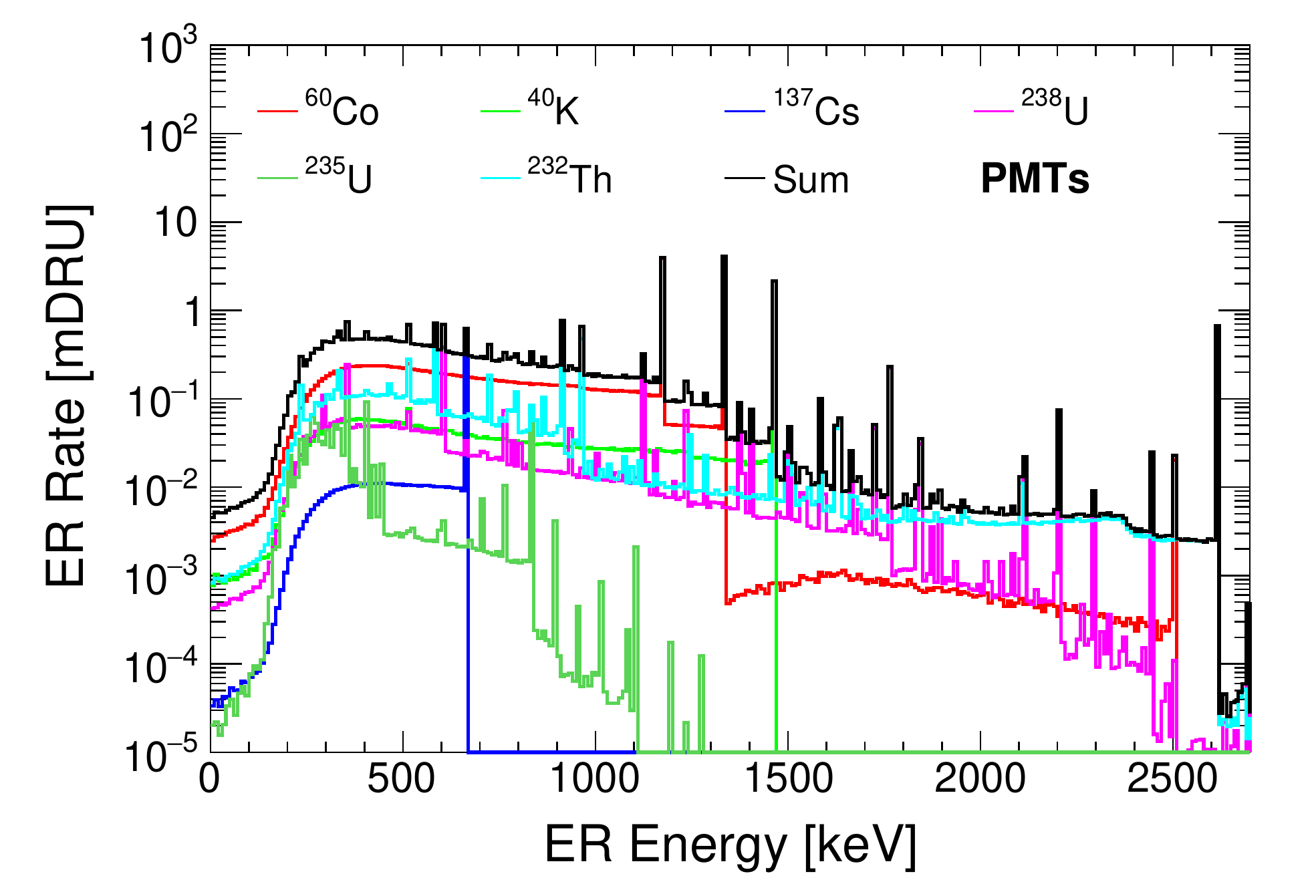}
  \caption{Electron recoil background from isotopes ($^{60}\rm{Co}$,
    $^{40}\rm{K}$, $^{137}\rm{Cs}$, $^{235}\rm{U}$, $^{238}\rm{U}$ and
    $^{232}\rm{Th}$). Results for the vessels and PMTs are plotted
    separately. }
  \label{er-isotopes}
\end{figure}

The ER rate spectrum is shown in Figure~\ref{er-spectrum}, from which
the background rates of various components can be determined. The
background rate is measured in mDRU ($=10^{-3}$ $\rm
events/day/kg/keV$). Various isotopes' contributions from the PMTs and
the SS vessels are shown in Figure~\ref{er-isotopes}. We
find most of the background comes from $^{232}\rm{Th}$, $^{238}\rm{U}$
and $^{60}\rm{Co}$.

The total ER background from the detector materials is estimated to be
0.021~mDRU, as summarized in Table~\ref{tab:er-material}. It indicates
that the largest contribution comes from the SS vessels due to their
large mass surrounding the TPC. The second largest contribution is
from the PMT arrays, as they are very close to the target volume.  A
20\% uncertainty is adopted based on the HPGe counting uncertainties.

\begin{table}[htbp] 
  \caption{ER background from detector materials for $E_{ee}$ within (1~keV, 10~keV).}
  \label{tab:er-material}
  \footnotesize
  \begin{tabular*}{80mm}{@{\extracolsep{\fill}}cc}
    \toprule\hline \multicolumn{2}{c}{\bf ER Background from materials} \\\hline
    Source &ER in mDRU   \\\hline
    Inner vessel      & 0.0064 $\pm$0.0013 \\\hline
    Outer vessel      & 0.0088 $\pm$0.0018 \\\hline
    PMT               & 0.0045 $\pm$0.0009 \\\hline
    PTFE              & 0.0005 $\pm$0.0001  \\\hline
    Copper            & 0.0008 $\pm$0.0002 \\\hline
    Total material    & 0.0210 $\pm$0.0042 \\\hline
    %\bottomrule
  \end{tabular*}
\end{table}

For internal background inside the liquid xenon, we first considered
$^{85}\rm{Kr}$. $^{85}\rm{Kr}$ $\beta$-decays with an end-point energy
687~keV and a half-life of 10.76 years. The abundance of
$^{85}\rm{Kr}$ in $^{Nat}\rm{Kr}$ has been measured as $2\cdot
10^{-11}$~\cite{kr85}. In PandaX-II, a 6.6$\pm$2.2~ppt of
$^{Nat}\rm{Kr}$ in xenon (mole/mole) had been achieved. In PandaX-4T,
a new online distillation tower will be constructed which is capable
of removing $^{Nat}\rm{Kr}$ level to 0.1~ppt.  The corresponding ER
background in the low energy region is estimated as 0.0053~mDRU. A
$20\%$ uncertainty is adopted due to the uncertainty in the shape of
the $\beta$ spectrum at low energies~\cite{kr85uncertainty}.

The contamination of dispersed $^{222}\rm{Rn}$ in the liquid xenon is
given particular attention. The $^{222}\rm{Rn}$ level measured in the
latest run of the PandaX-II is approximately
$8~\mu$Bq/kg~\cite{tan2016,cui2017} using delayed coincidence between
daughter $^{214}$Bi-$^{214}$Po. Most $^{222}\rm{Rn}$ atoms in the detector
come from the decay chain of $^{238}\rm{U}$. $^{222}\rm{Rn}$ can
emanate into target xenon through xenon plumbing, the cryostat and
detector components.
As the half-life (3.8 days) is relatively long, the distribution of
$^{222}\rm{Rn}$ is homogeneous in the target.
PandaX-4T is performing emanation measurement to screen the materials,
as well as implementing distillation and activated carbon filtration
systems. Our aim is to control the $^{222}\rm{Rn}$ level in xenon to
within $1~\mu$Bq/kg.  In the decay chain from $^{222}\rm{Rn}$ to the
stable $^{210}\rm{Pb}$, the $\beta$-decay of $^{214}\rm{Pb}$ dominates
the ER contribution. 1~$\mu$Bq/kg translates to an ER background of
0.0114~mDRU in the fiducial volume with a 10\% systematic uncertainty
of tagging $^{222}$Rn using $^{214}$Bi-$^{214}$Po coincidence.

The noble gas $^{220}\rm{Rn}$ has a lower ability to emanate into the
liquid xenon sensitive volume due to its shorter half-life. The
results from the PandaX-II experiment (see~\cite{tan2016,cui2017})
show that the background contribution from $^{220}\rm{Rn}$ is small
and is omitted here.

$^{136}\rm{Xe}$ is a two-neutrino double-beta decay isotope of xenon
with a half-life of $2.17\cdot 10^{21}$ years. The concentration of
$^{136}\rm{Xe}$ in natural xenon is approximately 8.9\%. In the
simulation, the double-beta decay energy spectrum is taken from the
DECAY0 code~\cite{decay0} and the average ER background rate in the
low energy region is found to be 0.0023~mDRU. A 15\% systematic
uncertainty is assumed, according to the discussions
in~\cite{xe136uncertainty}. {\color{black} The large amount of $^{136}\rm{Xe}$ in the sensitive volume
makes the PandaX-4T detector also sensitive to the search of neutrinoless double-beta decay~\cite{0vbb}.}

Electron neutrinos, $\nu_{e}$, from solar nuclear reactions,
contribute to the ER background at low energy by scattering
elastically off electrons in the liquid xenon. The dominant flux
component (92\%) of solar neutrinos is produced through the
proton-proton fusion with a neutrino energy up to 400 keV. An electron
capture reaction on $^{7}\rm{Be}$ is the second largest flux component
(7\%) in the low energy region, producing two mono-energetic neutrino
lines at 384.3 and 861.3~keV. The $pep$ and other sources contribute
to the remaining 1\%. In the simulation, we take into account the
neutrino oscillation and the reduced cross section for
$\nu_{\mu,\tau}$. Simulations show that the solar neutrinos contribute
an ER background of 0.009~mDRU, with a 2\% uncertainty covering the
systematic uncertainties of the neutrino flux and oscillation.

\begin{figure}[htbp]
  \centering
  \includegraphics[width=8cm]{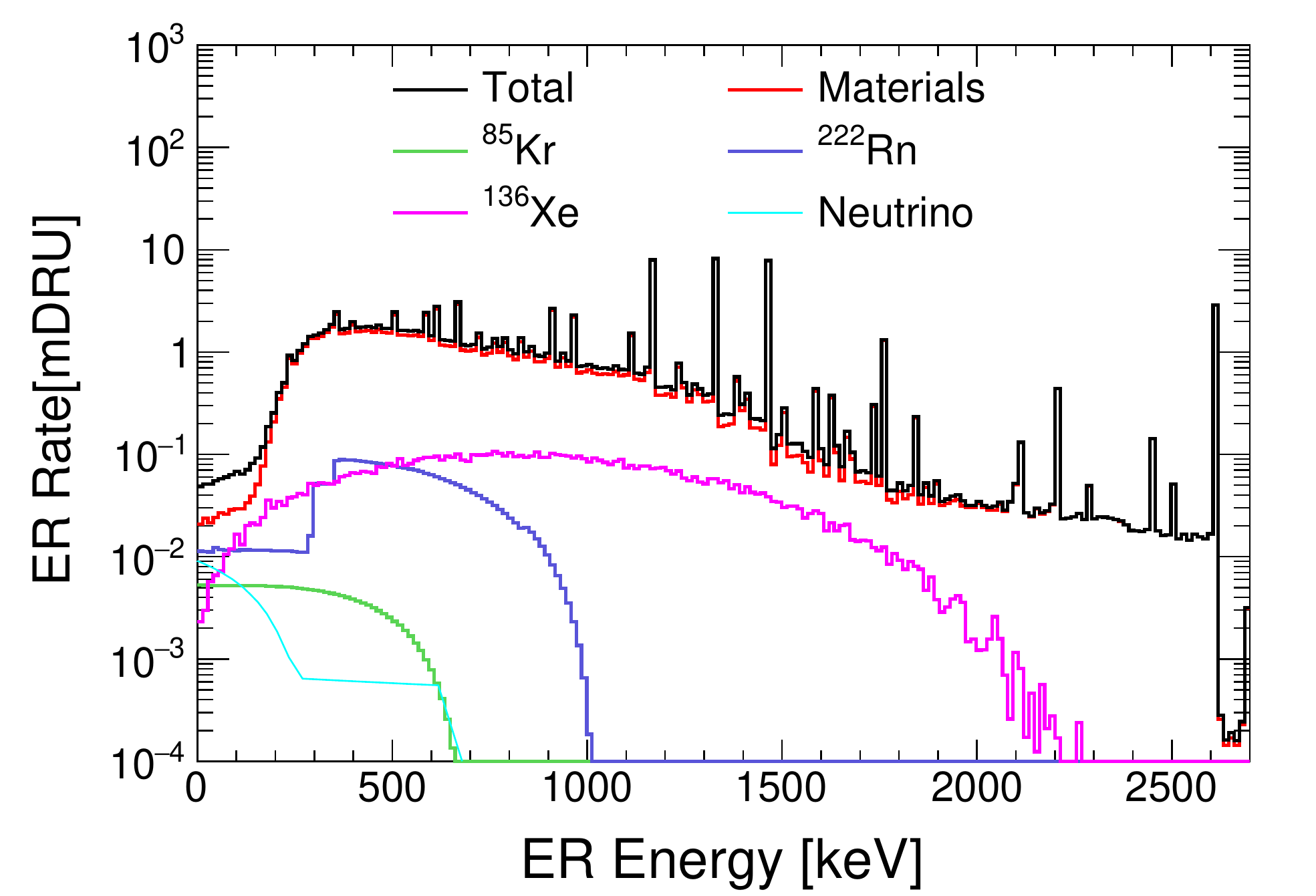}
  \caption{The ER background from detector materials, intrinsic contaminations and neutrinos. The bold black line is the total background.}
  \label{er-spectrum-all}
\end{figure}

Including all sources mentioned above, the full ER background spectrum
is shown in Figure~\ref{er-spectrum-all}. Based on the PandaX-II
analysis, we assume that such a background could be mostly rejected
using the $S2/S1$ ratio with a 99.75\% efficiency with 40\% NR
acceptance.

\subsection{NR Background}
\label{nr-background}
The NR background is produced by neutron background via elastic
scattering off the xenon nuclei.
As it is difficult to differentiate it from the WIMP-nucleus
scattering, it is crucial to control the neutron background.

The cosmogenic neutrons at CJPL has been suppressed to a negligible
level. Instead, the neutron background comes from primordial decay of
chains of $^{238}\rm{U}$, $^{235}\rm{U}$ and $^{232}\rm{Th}$ through
the spontaneous fission (S.F.) and through ($\alpha$,n) reactions. The
neutron yield from the ($\alpha$,n) reaction depends on the the
$\alpha$ rate in the primordial decay chains and the target
material. Common neutron producing targets include $^{13}\rm{C}$,
$^{14}\rm{N}$, $^{17}\rm{O}$, $^{18}\rm{O}$, $^{19}\rm{F}$, etc. For
PTFE, the neutron generation rate is high because of the abundance of
$^{19}\rm{F}$. For the neutron yield in PMTs, the ceramic stem ($\rm
Al_2 O_3$), the quartz window and the SS casing are simulated
separately. For heavy nuclei such as copper, the neutron generation
is mainly dominated by the S.F. from $^{238}\rm{U}$.

\begin{figure}[htbp]
  \centering
  \includegraphics[width=8cm]{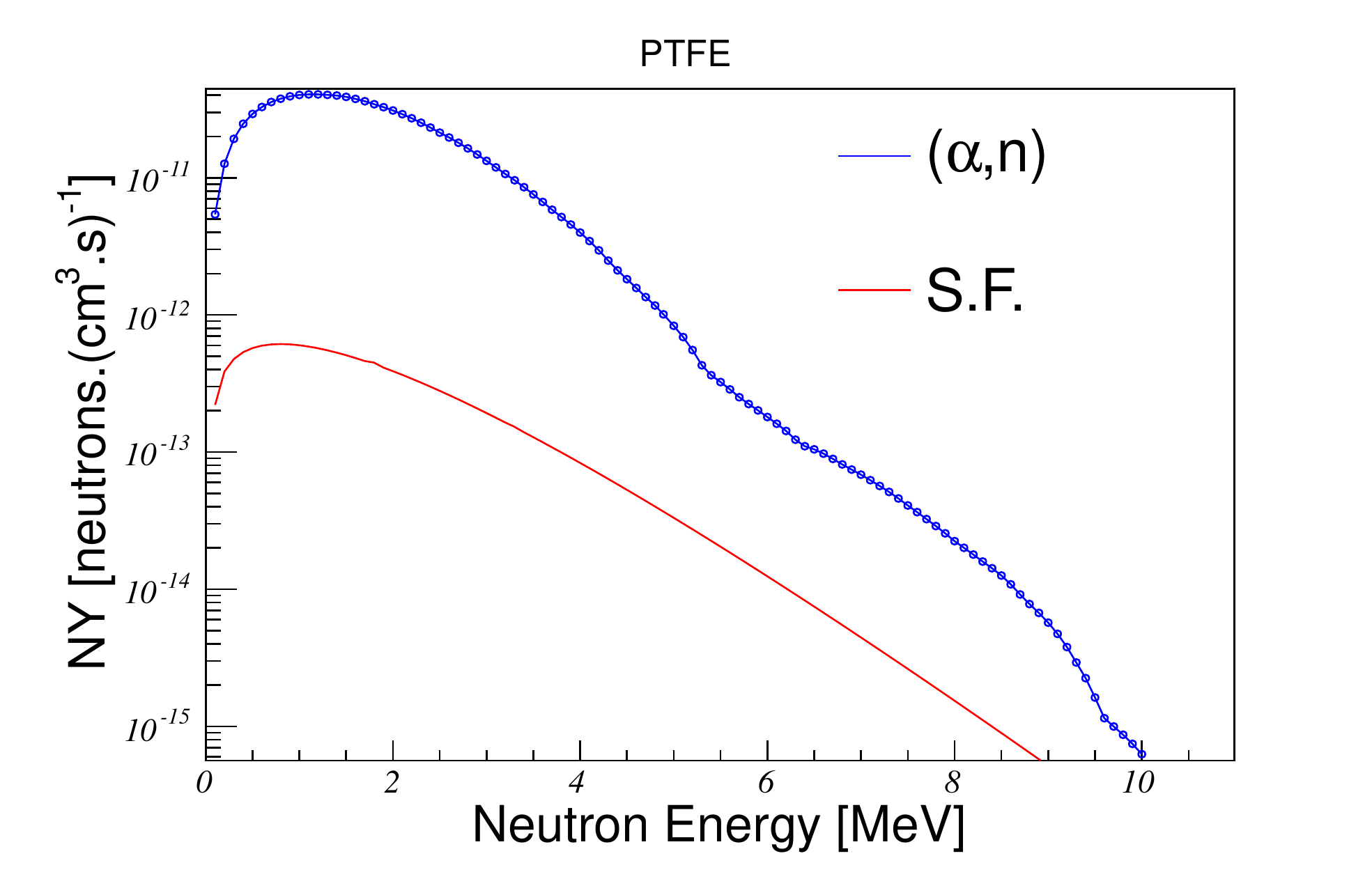}
  \includegraphics[width=8cm]{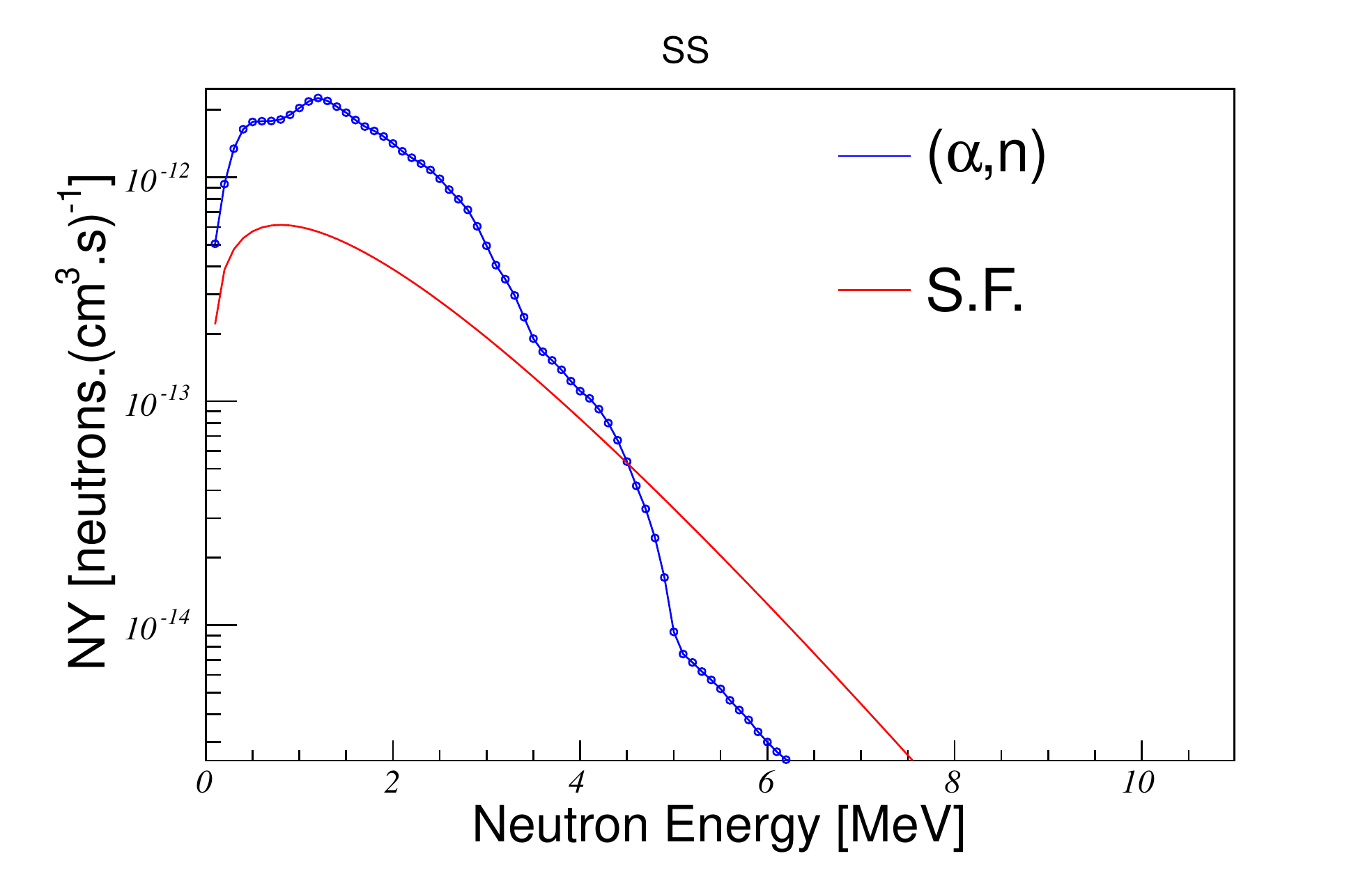}
  \caption{Neutron Yield: S.F. vs $(\alpha,n)$. Top: PTFE. Bottom: Stainless steel. It shows the fission dominates the neutron generation for stainless steel, while $(\alpha,n)$ mechanics dominates the PTFE. }
  \label{fission}
\end{figure}

With SOURCES4A~\cite{sources4a} package, we calculate the neutron
generation rates and energy spectra from S.F. and ($\alpha$,n)
mechanics in various materials, as shown in Figure~\ref{fission}.  In
all the materials, neutrons from the S.F. are dominated by the isotope
of $^{238}\rm{U}$ ($>99\%$). However, such a process 
emits more than one neutron or $\gamma$, which causes multi-scattering and can be
rejected in the selection~\cite{u238sf}. A similar effect exists in the ($\alpha$,n) 
reaction, that significant fraction of neutrons are emitted together with $\gamma$s, 
which may also be rejected. Such suppression is included in the Geant4 simulation below, 
but details will be described elsewhere~\cite{geant4neutron}.

\begin{figure}[htbp]
  \centering
  \includegraphics[width=8cm]{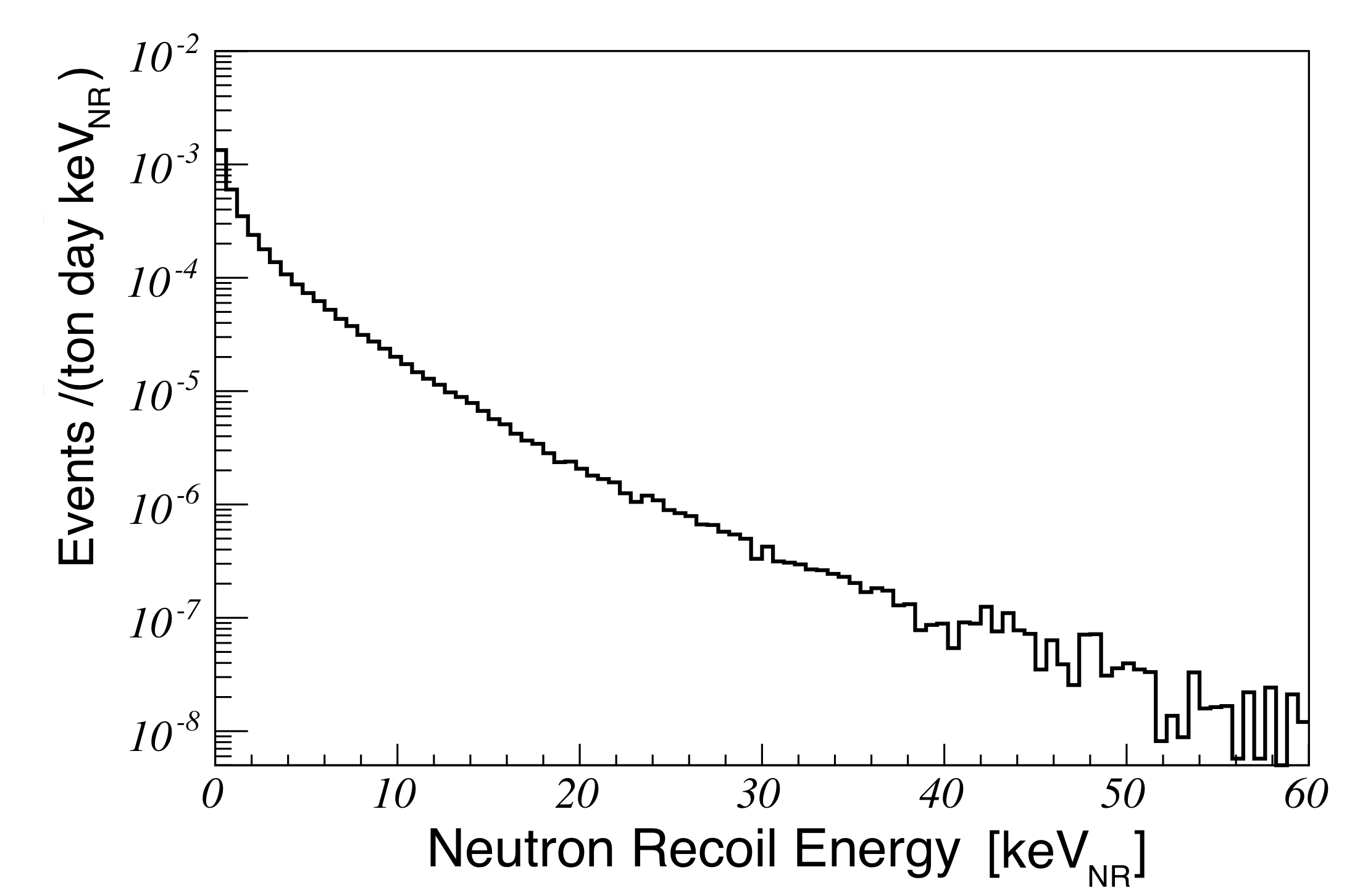}
  \caption{The NR background spectrum from simulation. }
  \label{nr-spectrum}
\end{figure}

\begin{figure}[htbp]
  \centering
  \includegraphics[width=8cm]{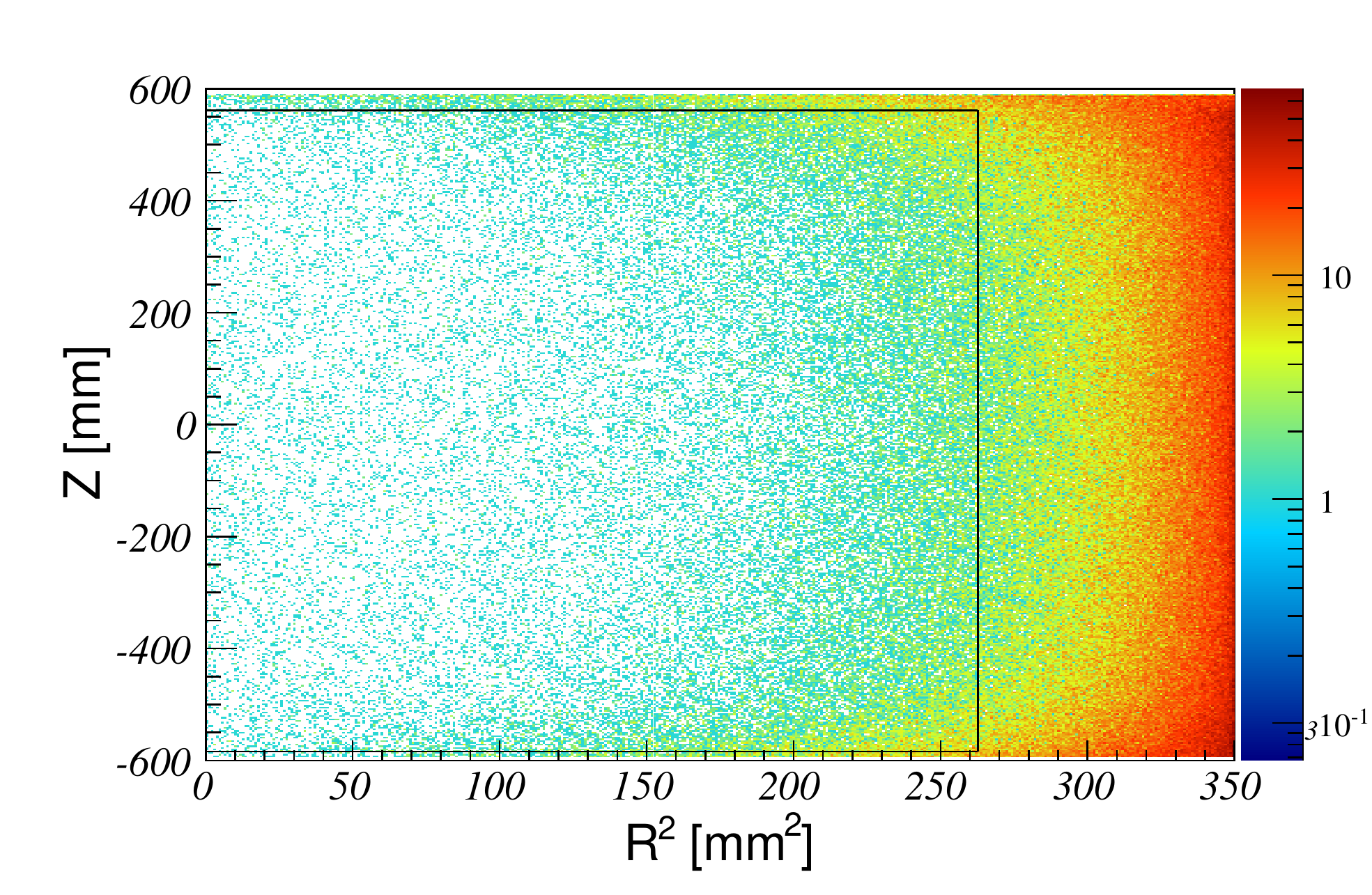}
  \caption{Distribution of NR background events in the target from simulation.}
  \label{nr-distribution}
\end{figure}

Using the BambooMC, we generate neutrons by sampling neutron energy
spectrum from SOURCE4A in the corresponding material, in combination
with $\gamma$ rays in the given ($\alpha$,n) reaction when
appropriate. We generate $10^6$ neutrons from each component and
simulate the energy deposition inside the liquid xenon. The obtained
NR energy spectrum and events distribution in the target are shown in
Figures~\ref{nr-spectrum} and \ref{nr-distribution}. The so-called
Lindhard factor is used in converting the NR energy into electron
equivalent energy~\cite{lindhard} when applying the energy selection
window. The simulation shows that the veto compartment can reject
approximately 50\% of the NR background from the TPC materials.

\begin{figure}[htbp]
  \centering
  \includegraphics[width=8cm]{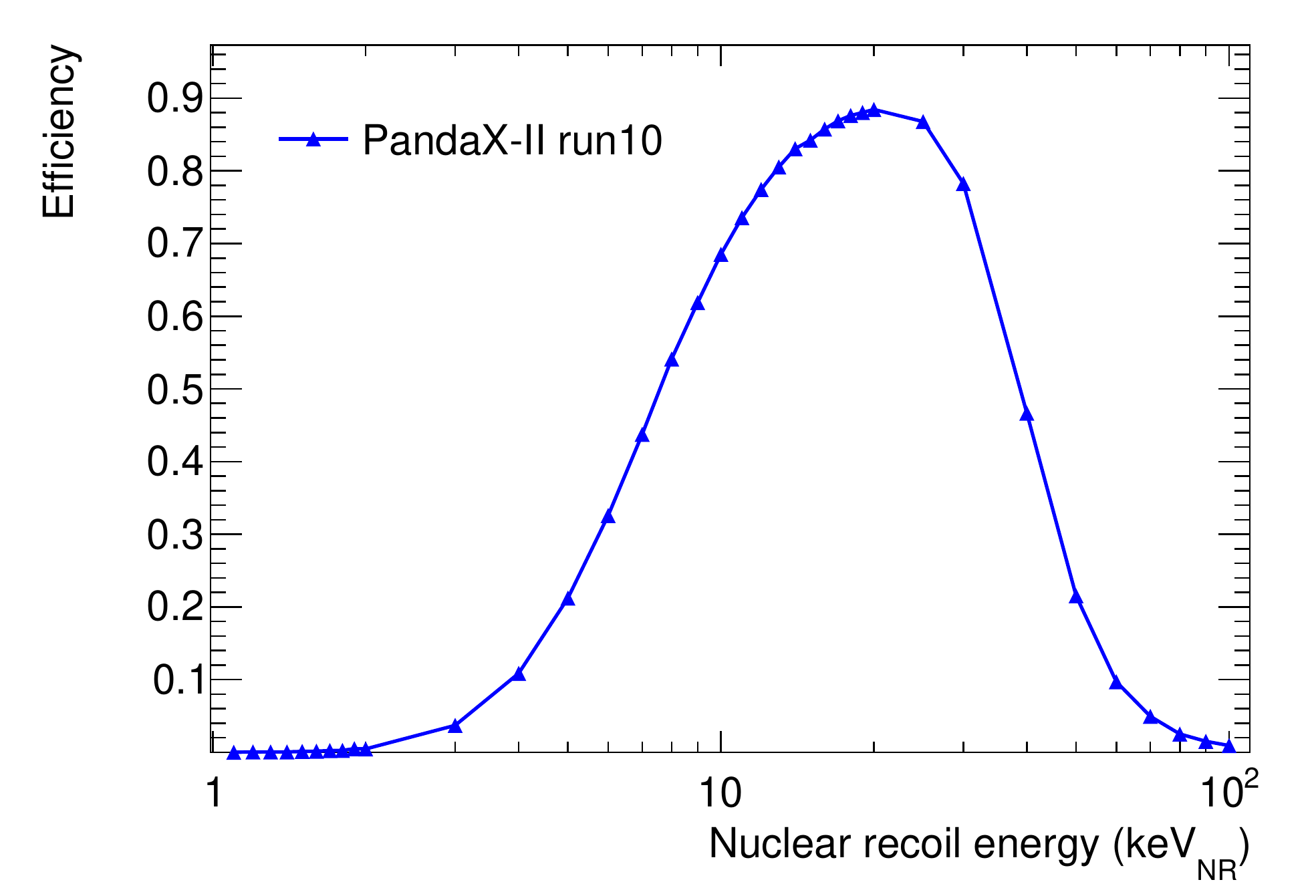}
  \caption{ NR events selection efficiency as a function of nuclear
    recoil energy in various runs of the PandaX-II.}
  \label{NReff}
\end{figure}

To calculate the detected rate of NR events, we apply the detection
efficiency as a funtion of NR energy obtained in the PandaX-II
experiment, as shown in Figure~\ref{NReff}.  Within the WIMP search
window, we predict the NR background from materials to be $2.0\cdot
10^{-4}$~mDRU, with detailed values from different components shown in
Table~\ref{tab:nr-material}. A 17\% systematic uncertainty is adopted
mainly from the neutron yield from the SOURCES4A.

\begin{table}[htbp] 
  \caption{NR background from detector materials for $E_{ee}$ within (1~keV, 10~keV).}
  \label{tab:nr-material}
  \footnotesize
  \begin{tabular*}{80mm}{@{\extracolsep{\fill}}cc}
    \toprule\hline \multicolumn{2}{c}{\bf NR Background from materials} \\\hline
    Source & NR rate in mDRU   \\\hline
    Inner vessel      & $4.6\pm 0.8 \cdot 10^{-5}$ \\\hline
    Outer vessel      & $5.3\pm 0.9 \cdot 10^{-5}$ \\\hline
    PMT Casing        & $3.5\pm 0.6 \cdot 10^{-5}$ \\\hline
    PMT Window        & $4.7\pm 0.8 \cdot 10^{-6}$ \\\hline
    PMT Stem            & $3.9\pm 0.7 \cdot 10^{-5}$ \\\hline
    PTFE              & $1.3\pm 0.2 \cdot 10^{-5}$ \\\hline
    Copper            & $5.9\pm 1.0 \cdot 10^{-6}$ \\\hline
    Total material    & $2.0\pm 0.3 \cdot 10^{-4}$ \\\hline
    %\bottomrule
  \end{tabular*}
\end{table}

\begin{figure}[htbp]
  \centering
  \includegraphics[width=8cm]{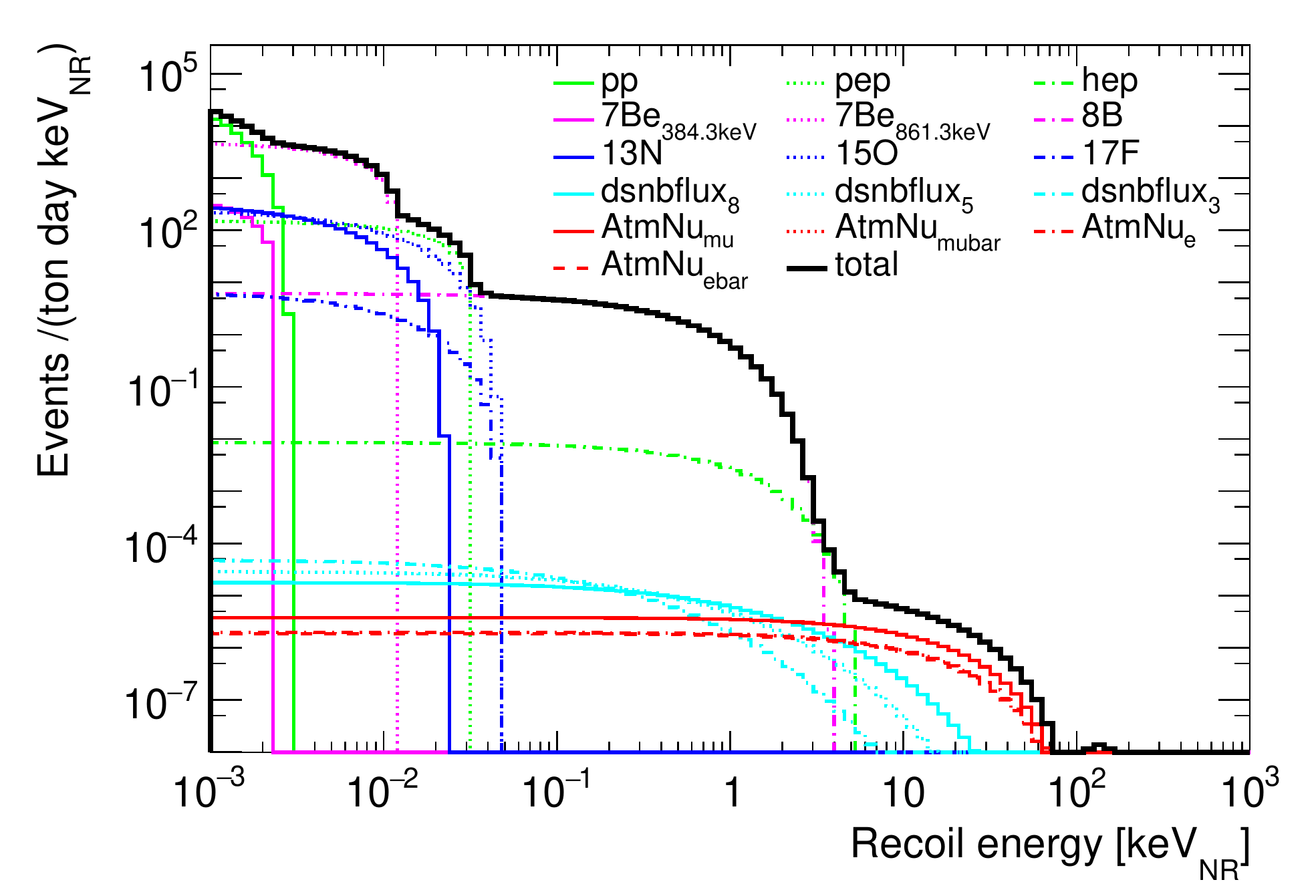}
  \caption{Event rate from coherent neutrino-nucleus scattering in xenon. The neutrio sources include solar neutrinos ($\rm ^{8}B$, $hep$, $\rm ^{7}Be$, CNO neutrinos, $pp$ chains), atmospheric neutrinos ($\nu_{e}$, $\bar{\nu}_{e}$, $\nu_{\mu}$, $\bar{\nu}_{\mu}$) and diffuse supernovae (${\rm T}_{\nu_{e}}$=3~MeV, ${\rm T}_{\bar{\nu}_{e}}$=5~MeV, ${\rm T}_{\nu_{x}}$=8~MeV).}
  \label{neutrinoNR}
\end{figure}

In addition to the radioactive isotopes, neutrinos can also generate
NR events through the coherent neutrino-nucleus scattering process
(CNNS), a process which was recently observed using accelerator
neutrinos by the COHERENT
Collaboration~\cite{neutrinocoherent}. The nuclear recoil spectra from various neutrino
sources are shown in Figure~\ref{neutrinoNR}.
For the dark matter candidate selection energy range, the dominant
contribution is from the $^8\rm B$ and $hep$ neutrinos.
Based on the PandaX-II efficiency, we estimate that the neutrino CNNS
contribution is $7.6\cdot10^{-5}$~mDRU with 50\% systematic
uncertainty due to selection efficiency and neutrino fluxes.

\subsection{The PandaX-4T Background Summary}
\label{background-summary}
As shown in Figure~\ref{er-dist} and Figure~\ref{nr-distribution}, the
ER and NR background events from the materials are concentrated close
to the edge of the target. A fiducial cut is chosen to suppress
background and to maintain a large effective mass: 35~mm below the
gate electrode, 10~mm above the cathode, and within 512.5~mm in the radial
direction. The total amount of xenon within this fiducial cut is
2.8~tons. Within this cut, the summary of ER and NR background is
given in Table~\ref{summary}. In the final background numbers, we have
also assumed that the final ER rejection efficiency is 99.75\% with
40\% NR acceptance. After a two-year exposure, the final expected background 
is 2.5 (ER) and 2.3 (NR) events. 
\begin{table}[htbp] 
  \caption{Final background budget within the WIMP search window.}
  \label{summary}
  \footnotesize
  \begin{tabular*}{80mm}{@{\extracolsep{\fill}}ccc}
    \toprule\hline \multicolumn{3}{c}{\bf Summary of ER and NR backgrounds} \\\hline
    Source &ER in mDRU & NR in mDRU  \\\hline
    Materials         & 0.0210$\pm$0.0042    & $2.0\pm 0.3 \cdot 10^{-4}$ \\\hline
    $^{222}\rm{Rn}$   & 0.0114$\pm$0.0012    &  -  \\\hline
    $^{85}\rm{Kr}$    & 0.0053$\pm$0.0011    &  -\\\hline
    $^{136}\rm{Xe}$   & 0.0023$\pm$0.0003    &  -\\\hline
    Neutrino          & 0.0090$\pm$0.0002    & $0.8\pm 0.4 \cdot 10^{-4}$\\\hline
    Sum               & 0.049 $\pm$0.005     & $2.8\pm 0.5 \cdot 10^{-4}$\\\hline
    2-year yield (evts)      & 1001.6$\pm$ 102.2    & 5.7$\pm$1.0  \\\hline
    after selection (evts) & 2.5$\pm$0.3            & 2.3$\pm$0.4 \\\hline
    %\bottomrule
  \end{tabular*}
\end{table}

\section{Physics Reach of the PandaX-4T}
\label{physics}

Given a WIMP mass and WIMP-nucleon scattering cross section, the NR 
rate and spectrum are calculated using identical formalism as in Refs.~\cite{cui2017}.
The NR efficiencies in PandaX-II
(Figure~\ref{NReff}) is adopted for the WIMP NR events as well.
For simplicity, we assume simple counting experiment with the
so-called $CLs$ method~\cite{cls}.

\begin{figure}[htbp]
  \centering
  \includegraphics[width=8cm]{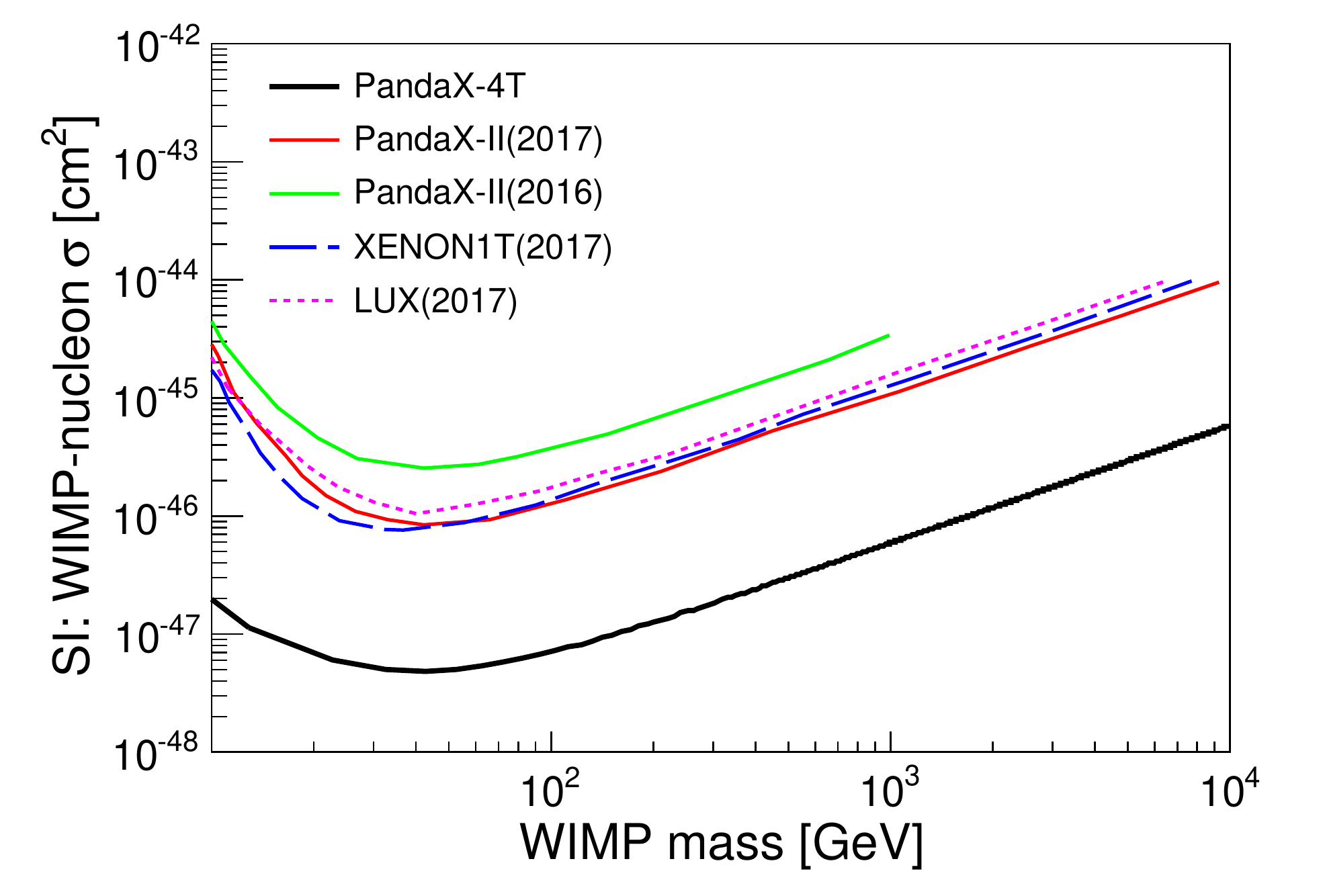}
  \caption{PandaX-4T sensitivity (at $90\%$ C.L.) to WIMP-nucleon
    recoil events with an exposure of 5.6~ton-year. We compare
    the result with other leading experiments: LUX~\cite{lux2017},
    XENON1T~\cite{xenon2017}, PandaX-II~\cite{tan2016,cui2017}. }
  \label{sen-si}
\end{figure}

\begin{figure}[htbp]
  \centering
  \includegraphics[width=8cm]{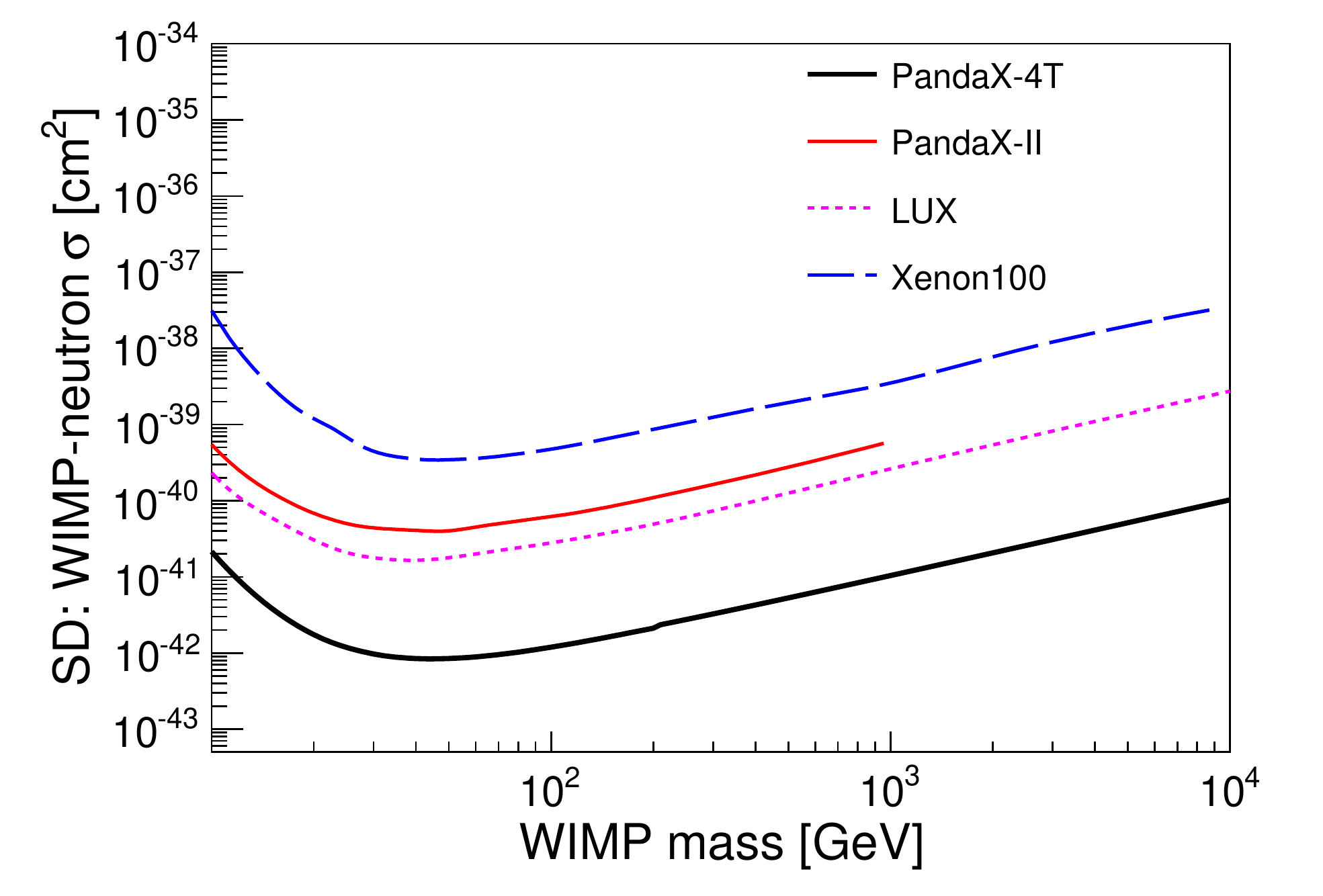}
  \includegraphics[width=8cm]{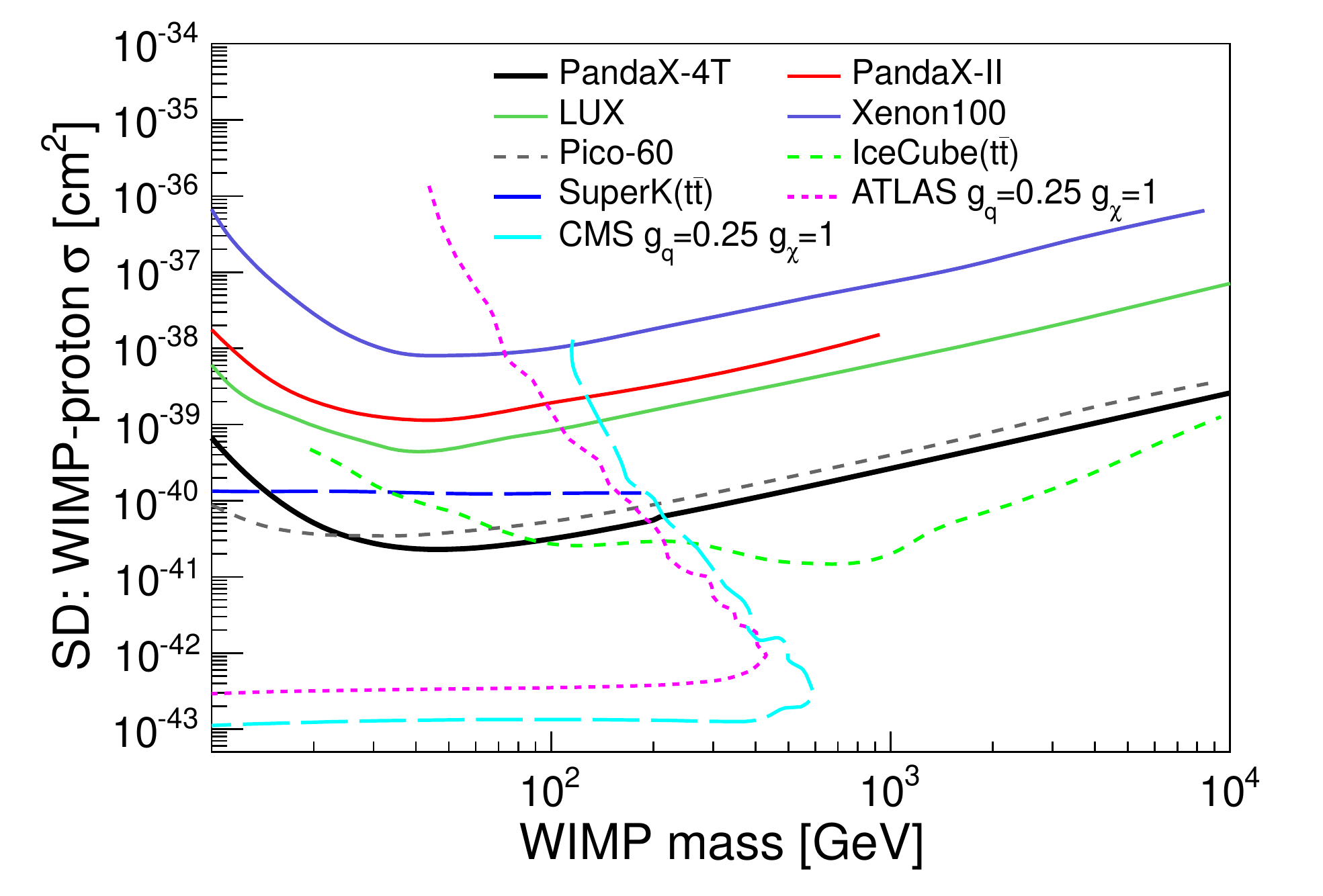}
  \caption{PandaX-4T sensitivity (at $90\%$ C.L.) to WIMP-nucleon
    recoil for spin-dependent interaction, upper: proton-only, lower:
    neutron-only.  We compare the results with the world data:
    LUX~\cite{luxSD}, XENON100~\cite{xenonSD},
    PandaX-II~\cite{fu2016}, CMS~\cite{monojetcms},
    ATLAS~\cite{monojetatlas}, PICO~\cite{pico1,pico2},
    IceCube~\cite{icecube} and Super-K~\cite{superk}. }
  \label{sen-sd}
\end{figure}

Under the expected background in
Table~\ref{summary},
the $90\%$ C.L. WIMP median sensitivity corresponds to a signal of 5 events after selection. The sensitivities for WIMP-nucleon spin-independent and
spin-dependent interactions are shown in Figure~\ref{sen-si} and
Figure~\ref{sen-sd}, respectively, with a 5.6-ton-year exposure.  For a
WIMP mass of $m_{\chi}=40~\rm GeV/c^2$, the sensitivity on the
interaction cross section reaches a minimum at $\rm 6\cdot
10^{-48}~cm^2$ for spin-independent interaction. For the
spin-dependent interaction, the strongest sensitivity reaches $\rm
9\cdot 10^{-43} cm^2$ for the WIMP-neutron-only coupling and $\rm
3\cdot 10^{-41} cm^2$ for the WIMP-proton-only coupling.

\section{Conclusion}
\label{conclusion}
PandaX-4T is a next generation dark matter direct detection experiment
with a multi-ton dual phase liquid xenon detector. In this paper we
present a comprehensive simulation study of the background from
radioactivity in the materials, intrinsic contaminations in the liquid
xenon and neutrinos through. The WIMP candidate selection is chosen to
be between 1~keV and 10~keV electron equivalent energy, single
scattering in anti-coincidence with the veto compartment, and a vertex
located in a 2.8-ton fiducial volume.  In the NR signal region (with
$S2/S1$ cut), we estimate the background to be 2.5$\pm$0.3 ER events
and 2.3$\pm$0.4 NR events for an exposure of 5.6 ton-year. The
expected WIMP sensitivity of PandaX-4T can be more than a factor of
ten improved from the currently running PandaX-II experiment. \\

\begin{acknowledgments}
  This project has been supported by a 985-III grant
  from Shanghai Jiao Tong University, grants from National Science
  Foundation of China (Nos. 11435008, 11455001, 11505112,
  11525522 and 11755001), and a grant from the Ministry of Science and Technology
  of China (No. 2016YFA0400301). We thank the support of grants from
  the Office of Science and Technology, Shanghai Municipal
  Government (No. 11DZ2260700, No. 16DZ2260200), and the support from
  the Key Laboratory for Particle Physics, Astrophysics and Cosmology,
  Ministry of Education. This work is supported in part by the Chinese
  Academy of Sciences Center for Excellence in Particle Physics
  (CCEPP) and Hongwen Foundation in Hong Kong. We also would like to
  thank Dr. Xunhua Yuan and Chunfa Yao of China Iron and Steel
  Research Institute Group, and Taiyuan Iron and Steel (Group) Co. LTD
  for crucial help on low background stainless steel. Finally, we
  thank the following organizations for indispensable logistics and
  other supports: the CJPL administration and the Yalong River
  Hydropower Development Company Ltd..
\end{acknowledgments}

%\Authorfootnote

%\end{multicols}

\begin{thebibliography}{90}
\vspace{3mm}
\bibitem{dm1}
  S. M. Faber and J. S. Gallagher, Ann. Rev. Astron. Astrophys. {\bf 17}, 135 (1979)
\bibitem{dm2}
  G.R. Blumenthal, S. Faber, J.R. Primack and M.J. Rees, Nature {\bf 311}, 517-525 (1984)
\bibitem{dm3}
  P. A. R. Ade {\it et al.} (Planck Collaboration), Astron. Astrophys. {\bf 594} (2016) A13 %arXiv:1502.01589
\bibitem{density}
  L. M. Widrow, B. Pym, J. Dubinski, Astrophys. J. {\bf 679}, 1239-1259 (2008)%, arXiv:0801.3414
\bibitem{wimp2}
  G. Bertone, D. Hooper and J. Silk, Phys. Rept. {\bf 405}, 279 (2005)
\bibitem{wimp3}
  D. Bauer {\it et al.}, FERMILAB-CONF-13-688-AE %arXiv:1310.8327
\bibitem{ddreview}
  J. Liu, X. Chen, X. Ji, Nature Phys. {\bf 13}, 212 (2017)
\bibitem{lux2017}
  D. S. Akerib {\it et al.} (LUX Collaboration), Phys. Rev. Lett. {\bf 118}, 021303 (2017) 
\bibitem{xenon2017}
  E. Aprile {\it et al.} (XENON1T Collaboration), Phys. Rev. Lett. {\bf 119}, 181301 (2017) %arXiv:1705.06655
\bibitem{tan2016}
  A. Tan {\it et al.} (PandaX Collaboration), Phys. Rev. Lett. {\bf 117}, 121303 (2016) 
\bibitem{cui2017}
  X. Cui {\it et al.} (PandaX Collaboration), Phys. Rev. Lett. {\bf 119}, 181302 (2017) %arXiv:1708.06917
\bibitem{xenon1t}
  E. Aprile {\it et al.} (XENON1T Collaboration), JCAP {\bf 04}, 027 (2016)
\bibitem{geant4}
  S. Agostinelli {\it et al.} (GEANT4 Collaboration), Nucl. Instrum. Meth. {\bf A 506} (2003) 250-303
\bibitem{kr85}
  X. Du, K. Bailey, Z.-T. Lu, P. Mueller, T.P. O'Connor and L. Young, Rev. Sci. Instr. {\bf 75}, 3223-3232 (2004)
\bibitem{kr85uncertainty}
  M. Selvi, in LOW RADIOACTIVITY TECHNIQUES 2013 (LRT 2013): Proceedings of the IV International Workshop in Low Radioactivity Techniques, vol. 1549, pp. 213-218, AIP Publishing, (2013)
\bibitem{decay0}
  O. Ponkratenko, V. Tretyak and Y. Zdesenko, Phys. Atom. Nucl. {\bf 63}, 1282-1287 (2000)
\bibitem{xe136uncertainty}
  J. Kotila and F. Iachello, Phys. Rev. C {\bf 85}, 034316 (2012)
\bibitem{0vbb}
  {\color{black} X. Chen {\it et al.}, Sci. China-Phys. Mech. Astron. {\bf 60}, 061011 (2017)}
\bibitem{sources4a}
  W. Wlison {\it et al.}, Tech. Rep., LA-13639-MS, Los Almos, 1999.
\bibitem{u238sf}
  S. Shaw, Ph.D. thesis, University of College London (2016)
\bibitem{geant4neutron}
  Q. Wang {\it et al.} (PandaX Collaboration), in preparation (2018)
\bibitem{lindhard}
  J. Lindhard {\it el al.}, Mat. Fys. Medd. Dan. Vid. Selsk., {\bf 33}, no. 10 (1963)
\bibitem{neutrinocoherent}
  D. Akimov {\it et al.} (COHERENT Collaboration), Science Vol. {\bf 357}, Issue 6356, 1123-1126 (2017)
\bibitem{cls}
  A. L. Read, J. Phys. {\bf G 28} (2002) 2693
\bibitem{luxSD}
  D. S. Akerib {\it et al.} (LUX Collaboration), Phys. Rev. Lett. {\bf 116}, 161302 (2016)%, arXiv:1602.03489
\bibitem{xenonSD}
  E. Aprile {\it et al.} (XENON100 Collaboration), Phys. Rev. D {\bf 94}, 122001 (2016)%, arXiv:1609.06154
\bibitem{fu2016}
  C. Fu {\it et al.} (PandaX Collaboration), Phys. Rev. Lett. {\bf 118}, 071301 (2017) 
\bibitem{monojetcms}
  A. M. Sirunyan {\it et al.} (CMS Collaboration), Phys. Rev. D {\bf 97}, 092005 (2018)
\bibitem{monojetatlas}
  M. Aaboud {\it et al.} (ATLAS Collaboration), J. High Energ. Phys. (2018) 2018:126 %arXiv:1711.03301
\bibitem{pico1}
  C. Amole {\it et al.} (PICO Collaboration), Phys. Rev. D {\bf 93}, 061101 (2016)%, arXiv:1601.03729
\bibitem{pico2}
  C. Amole {\it et al.} (PICO Collaboration), Phys. Rev. D {\bf 93}, 052014 (2016)%, arXiv:1510.07754
\bibitem{icecube}
  M. G. Aartsen {\it et al.} (IceCube Collaboration), JCAP {\bf 1604}, 022 (2016)%, arXiv:1601.00653
\bibitem{superk}
  K. Choi {\it et al.} (Super-Kamiokande Collaboration), Phys. Rev. Lett. {\bf 114}, 141301 (2015)

\end{thebibliography}
\end{document}